\documentclass[11pt,a4paper]{article}
\usepackage{float,mathtools}
\usepackage{graphicx}
\usepackage{amsmath}
\usepackage[bitstream-charter]{math design}
\usepackage[T1]{fontenc}
\usepackage{setspace,caption,hyperref}
\usepackage{undertilde}
\usepackage{lipsum}
\usepackage{subcaption,bbm}
\usepackage[left=2 cm,right=2 cm,top=2 cm,bottom=2 cm]{geometry}
\newtheorem{theorem}{Theorem}

\author{Anupam Kundu}
\begin{document}
\begin{center}
{\bf \huge Classification of Population Using Voronoi Area Based Density Estimation}\\
\end{center}

\begin{center}
{\bf \large Anupam Kundu\footnote{Texas A \& M University, Email: anupam.kundu@tamu.edu}, Subir Kumar Bhandari\footnote{Indian Statistical Institute, Kolkata, Email: subir@isical.ac.in}\\}
\end{center}
\doublespacing
\begin{quotation}
 
  Given two sets of training samples, general method is to estimate the density function and classify the test sample according to higher values of estimated densities. Natural way to estimate the density should be histogram tending to frequency curve. But nature of different amount of data accumulation at different points, obstruct us to take common interval length in drawing histogram. Here we followed a method similar to drawing histogram with an intention to density estimation and classification, but without doing arbitrarily we considered Voronoi diagram of the scattered plot and has obtained estimate of density based on areas of the Voronoi cells. But area of arbitrary Voronoi cell may be difficult to find out particularly in two and higher dimensions, and also in manifolds. For this reason we estimate the area of the Voronoi cell by drawing many independent uniform variates covering the support of the training samples. Please note that though we have in mind density estimation and classification problem in higher dimension and in typical manifolds, but our simulations and algebraic calculations have been mostly shown for densities on real line. We expect in higher dimension and in manifolds the work will be in similar lines and plan to do in near future. Also we expect discrete cases can be handled in slightly different but similar lines which are also left for future works.

\end{quotation}

\section*{Introduction}

Traditionally in Linear Discriminant Analysis (LDA), class-conditional probability densities are assumed to be normal with different mean vectors for different classes and identical covariance matrix. The resulting classification rule assigns an observation to the class for which the ratio of the normal densities at the point is greater than a constant. However sometimes it is not possible to assume the class dispersion matrices to be equal. Then the similar rule is stated as the Quadratic discriminant analysis. Anderson \cite{Anderson} provides an excellent introduction to discriminant analysis. \par 

 Discriminant analysis, linear as well as quadratic has generally proved to be highly effective in providing solutions to various classification problems. Despite being popular choice for the classification techniques the discriminant analysis do not perform well when the class conditional probability densities are very different from the normal distribution. There are some practical cases where we cannot have any idea of the parametric form the the class conditional densities from the training data set. So in order to handle the deviation from normality and better efficiency of the classifier, we often try to estimate the density function non-parametrically. \par 
 
 There are various methods for estimation of density functions non-parametrically in the existing literature. As for example we can state the most obvious density estimation process: the histogram. Apart from that there is an extremely popular method i.e. the Kernel density estimation method. The naive estimator, the nearest neighbour method, the variable kernel method, the orthogonal series estimator method etc. are also some popular choices of density estimation procedure. Silverman \cite{Silverman} describes all these density estimation procedures with special focus on Kernel density estimation method. The use of Kernel density estimates for the Bayes classification technique is quite popular among the scientists and engineers. The idea of nearest neighbour density estimate was presented in Fix and Hodges\cite{Fix}. In Loftsgaarden and Quesenberry\cite{Loft} a nearest neighbour method with the number of nearest neighbour as a function of the number of observation is considered for density estimation and its consistency has been studied.  Ghosh, Chaudhuri and Sengupta \cite{Ghosh1}(2006) presents a multi-scale(different scales of smoothing of the kernel density estimates) approach along with a graphical device leading to a more informative discriminant analysis. In the above mentioned paper a p-value type discriminant measure is studied. In Ghosh and Chaudhuri\cite{Ghosh2}(2004), a critical investigation is made into the influence of the value of the bandwidth on the behaviour of the average misclassification probability of a kernel based classifier. In the book by Biau and Devroye \cite{Gerard} properties of the k-nearest neighbour density estimates, the convergence with respect to a Lebesgue measure etc. are studied.

\section*{Proposed Method}


  In our case we have used the idea of histogram and Voronoi diagram (described in the book by Preparata and Shamos \cite{Shamos}) to do the density estimate and observed how this estimate behaves in terms of misclassification probability. In the next part we will describe the most simple case in our density estimation process i.e. the Voronoi area based nearest neighbour estimation method.\par 
  
  Suppose we have a training set from two distributions $f_1(x)$ and $f_2(x)$ in one dimension i.e. $X_{11},X_{12},$ $\dots X_{1{n_1}}$ and $X_{21},X_{22},\dots X_{2{n_2}}$ from training set 1 and 2 respectively. We want to estimate the distributions corresponding the training sets by using $X_{ij}$'s. We first denote the highest and the lowest observations from the training set and calculate $K_i=\max{(|X_{i(n_1)}|,|X_{i(1)}|)}$ where $X_{i(n_1)}$ and $X_{i(1)}$ are the highest and lowest order statistics of the i-th ($i=1,2$) training sets. In order to estimate the densities we first generate two sets of uniform distributions as follows: $U_i \sim Unif(-K_i,K_i)$. We generate $U_{i1},U_{i2},\dots U_{iN_i}$ i.e. $N_i$ observations from each uniform distribution. These $U_{ij}$'s will be used to estimate the area of the cells in Voronoi Diagram of the training set observations in the $i-th$ class. We should generate uniforms having support on a set which covers the major part of the support of the distribution. For example in case of normal we will pick uniforms from the $3\sigma$ bounds on both sides of the mean. So not always we need to use the $K_i$ boundary for support of the uniform distribution. However this can be a fair choice for densities with infinite support. For densities with finite support we should choose the support of the underlying uniform in a way such that it covers the support of the density. Slightly big support is helpful because it will help us to estimate the large Voronoi cell area at the boundary of the support of the distribution.  For example if we want to estimate the density of $Uniform(0,1)$, we may use $U(-1,2)$ for estimation of the Voronoi cell. So let us assume that $U_i\sim(l_{i1},l_{i2})$ for suitable $l_{i1}$ and $l_{i2}$.  Note that bigger or smaller values of $l_{i1}$ and $l_{i2}$ will not affect our density estimation by method using Voronoi diagram as will be clear from the subsequent discussions. But the interval should contain the approximate support which is necessary. Now suppose we define $$g_{ij}=\text{Number of $U_{ik}$'s that are nearest to $X_{ij}$, where $i=1,2$ and $j=1,2 \dots n_i$ }$$ As we can see from the definition of the $g_{ij}$'s that $\sum_{j=1}^{ni} g_{ij} = N_i$. Now we will use these $g_{ij}$'s
as corresponding weights. Let $w_{ij}= \frac{g_{ij}}{N_i}$ where $i=1,2$ and $j=1,2$ \dots $n_i$ will be proportional to the area of the Voronoi cells. From the definition of $w_{ij}$'s, it is clear that $$\sum_{j=1}^{ni} w_{ij} = 1 \qquad \sum_{j=1}^{ni} w_{ij} \hat{f}(X_{ij}) = 1 \qquad \hat{f}(X_{ij}) \propto \frac{1}{w_{ij}}$$ So using these we get the estimate of the density function. Suppose we want to estimate the density at the point $x$. Then first we find the distance from $x$ to all $X_{ij}$'s. Suppose the minimum distance occurs at $i_0$ for the first class and $i_1$ for the second class. Then the corresponding density estimates will be as follows:
$$\hat{f_1}(x)= \frac{N_1}{g_{1i_0}n_1}\qquad \hat{f_2}(x)= \frac{N_2}{g_{2i_1}n_2}$$ Now we want to see the behaviour of this density function estimate and how it performs in Bayes classification method  and also asymptotically.\par 

   As found by the theoretical discussion, estimate of density based on one Voronoi Cell has expectation close to the value of the density at that point but with some positive variance, shown in the subsequent result. So it is expected as also seen from the simulations, the above method based on one Voronoi cell will perform moderately well for the estimation of density and also for the classification problem. But we can improve by clubbing some neighbouring Voronoi cells, stated by us as Nearest Neighbour method so as to decrease variance of the density estimate and also for better classification.  

  Now this method works fine when the densities have full support of the real line but fails to work when the support of the density is on a particular set or union of some sets. This is because in this method we are trying to calculate the nearest neighbour of the point for which the density is going to be estimated and use the corresponding weight  to estimate the density at that particular point.While doing this even if the point falls outside the support of the density function, it can still find a nearest neighbour and use that corresponding weight, providing a false estimate of the density function. For example we tried with $U(0,1)$ and found out that it provides the estimate close to 1 outside the interval $(0,1)$.  In order to deal with this anomalous behaviour, we modified this method to address this problem.  \par 
  
  So in the set up as above we have done some modifications.  We have previously defined the frequencies of uniforms taken to estimate the Voronoi area in 1 dimension. Our idea of modification is that the Voronoi cell at the boundary of the support has a very large area. As the frequency of the uniforms will be proportional to the area of the cell and the density is inversely proportional to the corresponding frequency, so the density of the point outside the support will be close to zero.  In the next paragraph we will describe the method for Voronoi Area Based k Nearest Neighbour Density Estimation Method which is simultaneously a modification and generalisation of the method described above.\par 
  
   We have already defined $g_{ij}$'s. Now we define another quantity : 
   \begin{align*}
   g_{ij}^{*}&= \text{sum of the frequencies of the nearest neighbours of $X_{ij}$ if $\left(\frac{g_{ij}}{\sum_j g_{ij}}\right)< r $}\\
   &\text{ where $r$ is a previously fixed small quantity. for $i=1,2$ and $j=1,2,\dots n_i$}\\
   &= k*g_{ij} \qquad\text{  if $\left(\frac{g_{ij}}{\sum_j g_{ij}}\right)\geq r$ }\\ 
   g_{ij}^{**}&=\left(l_{i2}-l_{i1}\right) \times \frac{g_{ij}^*}{\sum_j g_{ij}^*} \qquad \text{These are the estimate of the areas of the Voronoi cells.}
   \end{align*}
    From this our new estimate of the density becomes the following:
    $$\hat{f}_{1}(x)=\frac{1}{n_1 g_{1j}^{**}}$$  where $g_{1j}^{**}$ is the estimate of the area of the Voronoi cell in which $x$ lies i.e. $x$ is nearest to $X_{1j}$.\par 
    
    Here we consider asymptotic properties of the density estimate using estimate of the Voronoi cells and the corresponding nearest neighbour method. Discussions are done for real line and we expect for support in general space the theory can be thought out. For calculations of the asymptotic-s, we develop the following results in terms of order statistics. The relation of our density estimate with the results derived is obvious. \par 
    
\section*{Theoritical Work}

\begin{theorem}
\label{Th1}
Let $X_{(1)}\leq X_{(2)}\leq \dots \leq X_{(n)}$ are order statistics of $X_1,X_2,\dots X_n$. Let us define 
$w_i=\frac{1}{2}[X_{(i+1)}-X_{(i-1)}]$. We have $$\mathbb{E}\left[\frac{n}{2}[X_{(i+1)}-X_{(i-1)}]\right]\to \frac{1}{f(x_0)} \qquad \text{as $n\to\infty$ and $\frac{i}{n} \to q$, where $X_1, X_2\dots \overset{iid}{\sim} F$} $$ $x_0$ being $q-th$ quantile of $F$. Obviously $i\to \infty$ as $n\to\infty$.  $f$ is continuous density of $F$ and $f(x_0)>0$.  
\end{theorem}

 \subsection*{Discussion and Proof of Theorem~\ref{Th1}}

Use the  transformation $U_i=F(X_i)$. $U_{(1)},\dots U_{(n)}$ are the order statistics of the $Uniform(0,1)$ distribution. $P[U_{(n)}<x]=x^n$ and therefore $\mathbb{E}[U_{max}]=\frac{n}{n+1}$  and  $P[U_{(1)}<x]=1-(1-x)^n$ and therefore $\mathbb{E}[U_{min}]=\frac{1}{n+1}$. We know the distribution of the i-th order statistic is the following:
$$f(U_{(i)}=x)=\frac{n!}{(i-1)!(n-i)!} F^{i-1}(x)f(x)(1-F(x))^{n-i}$$ Therefore from this we can calculate $\mathbb{E}[U_{(i)}]$. 
\begin{align}
\mathbb{E}[U_{(i)}]&=\frac{n!}{(i-1)!(n-i)!}\int_{0}^{1}x.x^{i-1}(1-x)^{n-i}dx \nonumber\\
&=\frac{n!}{(i-1)!(n-i)!} Beta(i+1, n-i+1)  \nonumber\\
&=\frac{i}{n+1} \qquad \text{by using this result:  $Beta(\alpha, \beta)=\frac{\Gamma(\alpha)\Gamma(\beta)}{\Gamma(\alpha+\beta)}$ and $\Gamma(r)=(r-1)!$}
\end{align} 
 From this result we can show that $\frac{n}{2}[U_{(i+1)}-U_{(i-1)}]\to 1$. This is because:
 \begin{align*}
 \mathbb{E}\left[\frac{n}{2}[U_{(i+1)}-U_{(i-1)}]\right]&=\frac{n}{2}\mathbb{E}[U_{(i+1)}-U_{(i-1)}]\\
 &=\frac{n}{n+1} \to 1 \qquad  \text{as $n\to \infty$}
 \end{align*}
 Next we want to find out the variance of the above quantity i.e. $Var\left[\frac{n}{2}[U_{(i+1)}-U_{(i-1)}]\right]$ In order to calculate the variance we need to calculate $\mathbb{E}[U_{(i)}^2]$ and $\mathbb{E}[U_{(i)}U_{(j)}]$
  The joint probability density function of uniform order statistics $U_{(i)}$ and $U_{(j)}$ where $i<j$ is the following:
  $$f_{U_{(i)},U_{(j)}}(u,v)=\frac{n!}{(i-1)!(j-i-1)!(n-j)!} u^{i-1} (v-u)^{j-i-1}(1-v)^{n-j}$$
 Using this we can easily calculate the $\mathbb{E}[U_{(i)}U_{(j)}]=\frac{i(j+1)}{(n+1)(n+2)}$. 
 \begin{align}
  \mathbb{E}[U_{(i)}U_{(j)}]&=\frac{n!}{(i-1)!(j-i-1)!(n-j)!} \int_{0}^{1}\int_{0}^{y} yx^{i}(y-x)^{j-i-1}(1-y)^{n-j}dx dy \nonumber\\
  &=\frac{n!}{(i-1)!(j-i-1)!(n-j)!}  Beta[i+1,j-i]\int_{0}^{1} y^{j+1}(1-y)^{n-j}dy \nonumber\\
  &= \frac{n!}{(i-1)!(j-i-1)!(n-j)!} Beta[i+1,j-i] Beta[j+2,n-j+1] \nonumber\\
  &=\frac{i(j+1)}{(n+1)(n+2)}
 \end{align}
 
If we put $i=j=r$ the we get, $\mathbb{E}[U_{(r)}^2]=\frac{r(r+1)}{(n+1)(n+2)}$ , So 
\begin{align}
\mathbb{E}\left[\frac{n}{2}\left(U_{(i+1)}-U_{(i-1)}\right)\right]^2&=\frac{n^2}{4}\mathbb{E}\left[U_{(i+1)}^2+U_{(i-1)}^2-2U_{(i+1)}U_{(i-1)}\right]\nonumber\\
&= \frac{6 n^2}{4(n+1)(n+2)}\\
Var\left[\frac{n}{2}\left(U_{(i+1)}-U_{(i-1)}\right)\right]&=\frac{n^2(n-1)}{2(n+1)^2(n+2)} \to \frac{1}{2}  \qquad \text{as $n \to \infty$}
\end{align}
  We shall find estimator of $\frac{1}{f(x_0)}$ from the uniform distribution. $T=\frac{n}{2}\left[ U_{(i+1)}-U_{(i-1)}\right]$ where $\mathbb{E}[T]=\frac{n}{n+1} \to 1$ and $Var[T]=\frac{n^2(n-1)}{2(n+1)^2(n+2)} \to \frac{1}{2}$. Therefore  this estimator $T$ of density of uniform is asymptotically unbiased but has asymptotic standard deviation as $\frac{1}{\sqrt{2}}$. 
  
  Similarly if $T_{r,m}=\frac{n}{m}\left[ U_{(r+m)}-U_{(r)}\right]$, then also $\mathbb{E}[T_{r,m}]=\frac{n}{n+1} \to 1$ and $Var[T_{r,m}]=\frac{n^2(n-m+1)}{m(n+1)^2(n+2)} \to \frac{1}{m}$ as $n\to \infty$, provided m is fixed. This expectation and variance both are free of the quantity $r$.\par
  
  \begin{align}
  \label{Eq1}
  \lim_{b\to a}\frac{F^{-1}(a)-F^{-1}(b)}{a-b}=\lim_{q\to p}\frac{p-q}{F(p)-F(q)} \to \frac{1}{f(x)}\qquad \text{if $x\in (b,a)$}
  \end{align}
  
  where $F^{-1}(a)=p$ and $F^{-1}(b)=q$.  Now suppose $x_0\in \left( U_{(i-1)},U_{(i+1)}\right)$.  
  \begin{align}
  \mathbb{E}\left[\frac{n}{2}\left[X_{(i+1)}-X_{(i-1)}\right]\right]
 &= \mathbb{E}\left[\frac{n}{2}\left[U_{(i+1)}-U_{(i-1)}\right]\times \frac{F^{-1}\left(U_{(i+1)}\right)-F^{-1}\left(U_{(i-1)}\right)}{U_{(i+1)}-U_{(i-1)}}\right] \nonumber\\
 & \to 1.\frac{1}{f(x_0)} = \frac{1}{f(x_0)}
  \end{align}
 This is because the first term of the product i.e. $T$ converges weakly to 1. Hence it also converges in probability. The second term also converges weakly to a constant. Therefore by Slutsky's Theorem $$nw_i\to \frac{1}{f(x_0)}$$ Using ~\ref{Eq1} and very similar  equations we have the following theorems.
 
 \begin{theorem}
 $Var\left[\frac{n}{2}\left(X_{(i+1)}-X_{(i-1)}\right)\right] \to \frac{1}{2f^2(x_0)}$
 \end{theorem}
  
\begin{theorem}
If $S_{r,m}=\frac{n}{m}\left[ X_{(r+m)}-X_{(r)}\right]$, then also $\mathbb{E}[S_{r,m}]\to \frac{1}{f(x_0)}$ and $Var[S_{r,m}]\to \frac{1}{mf^2(x_0)}$ provided $m$ is fixed.
\end{theorem}

\subsubsection*{Comment}
 We take $m$ big but $\frac{m}{n}\to 0$. As for example we may take $m=\sqrt{n}$ etc. If $m$ is big , variance of $S_{r,m}$ being proportional to $\frac{1}{m}$ become negligibly small.  Hence $S_{r,m}$ gives a good estimate of $\frac{1}{f(x_0)}$, where $n\to \infty $ such that $\frac{r}{n}\to constant$. \par 
 
 Here from asymptotic properties of $S_{r,2}$, we get properties of one nearest neighbour density estimate using one Voronoi cell, which is asymptotically unbiased (using reciprocal). For one nearest neighbour, the variance of the density estimate is positive. Now Using $S_{r,m}$ we approximate $(m-1)$ nearest neighbour density estimate using Voronoi cell. In fact we are clubbing different Voronoi cells to decrease the variance of the density estimate (which is asymptotically unbiased). If $m$ is big the variance becomes small.

 \subsection*{Comment}
  In case of K-nearest neighbour if we take k-nearest neighbour of some point outside the support, it may include $(k-1)$ points (out of $k$ points) inside the support which will affect the estimate of the density. For these reasons we have introduced a a small constant $r$ and if the relative frequency of the cell exceeds $r$, we expect the points of the cell is outside the support. In this case we consider all its nearest neighbours to be itself, which in turn increases the weight of the cell making the density at this point negligibly small.

  \section*{Simulation}
  
  We have used this method in one dimensional set-up and the corresponding misclassification probability while using the estimated density for Bayes classification. We have used our method for simulated data. Here we are going to present 5 one dimensional density comparison along with two 2-dimensional density comparison (one on circle) and how our method estimates the density and classify the observation using the ratio of the estimated density by Bayes classification rule.
  
 \subsection*{ Case 1}

\begin{figure}
\centering
\begin{subfigure}{.45\textwidth}
  \centering
  \includegraphics[width=.9\linewidth]{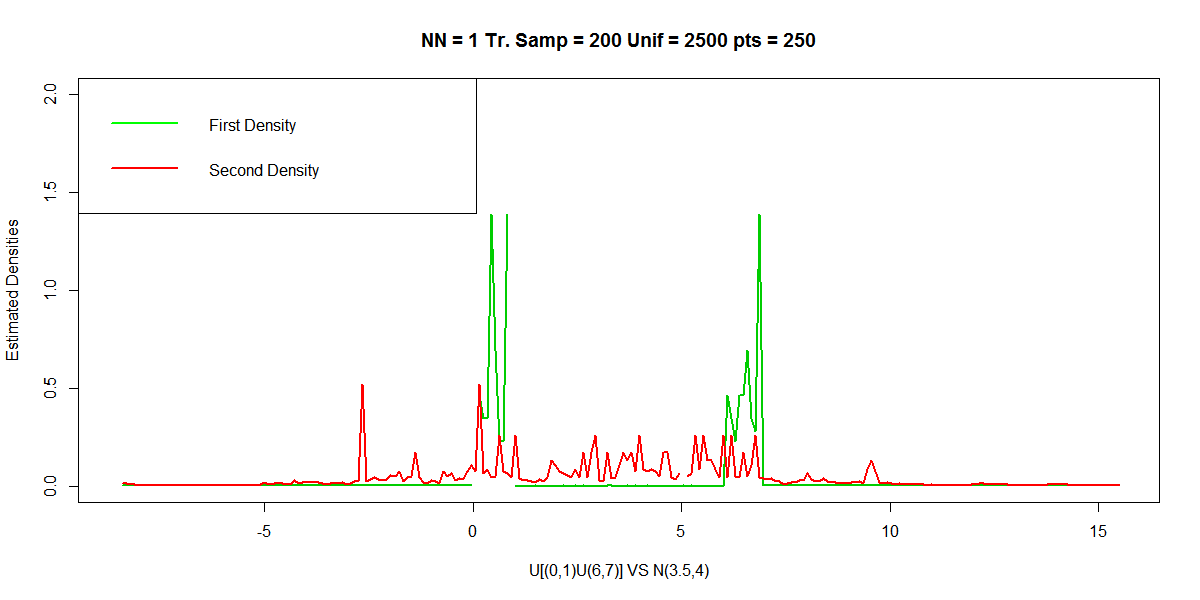}
  \caption{mUN-TS200-NN1}
  \label{mUN-TS200-NN1}
\end{subfigure}
\begin{subfigure}{.45\textwidth}
  \centering
  \includegraphics[width=.9\linewidth]{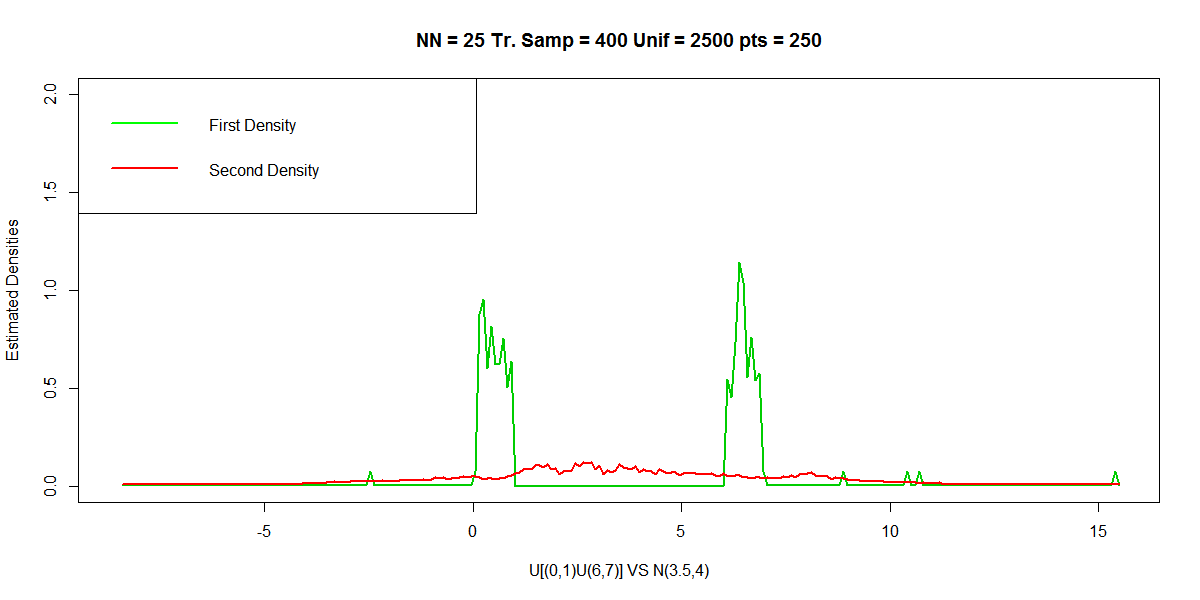}
  \caption{mUN-TS200-NN10}
  \label{mUN-TS200-NN10}
\end{subfigure}
\begin{subfigure}{.45\textwidth}
  \centering
  \includegraphics[width=.9\linewidth]{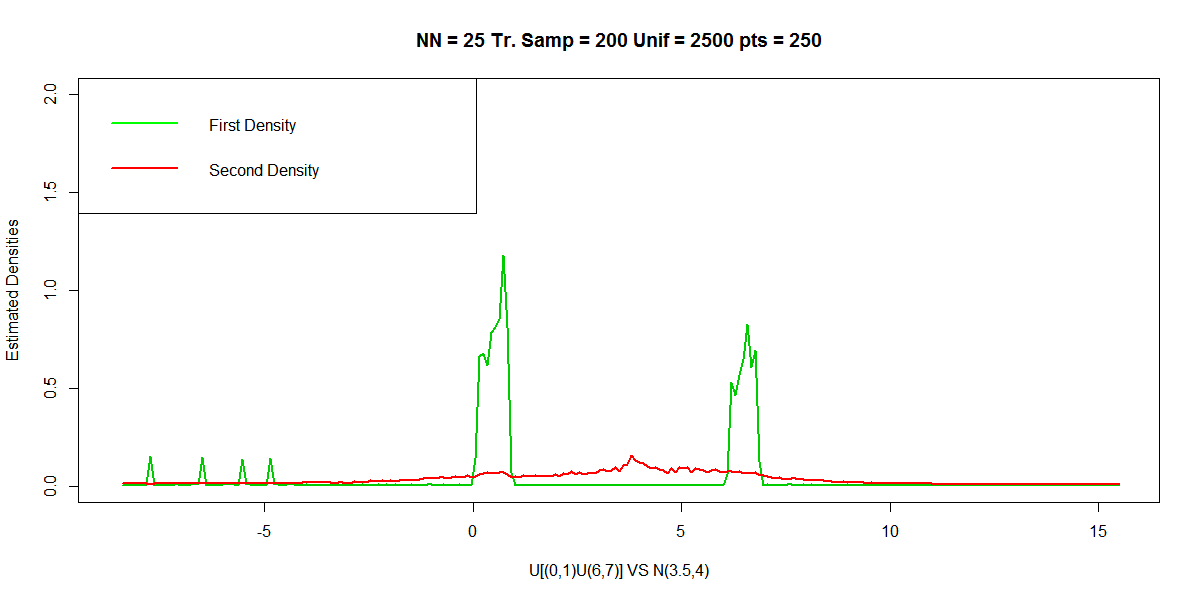}
  \caption{mUN-TS200-NN25}
  \label{mUN-TS200-NN25}
\end{subfigure}
\begin{subfigure}{.45\textwidth}
  \centering
  \includegraphics[width=.9\linewidth]{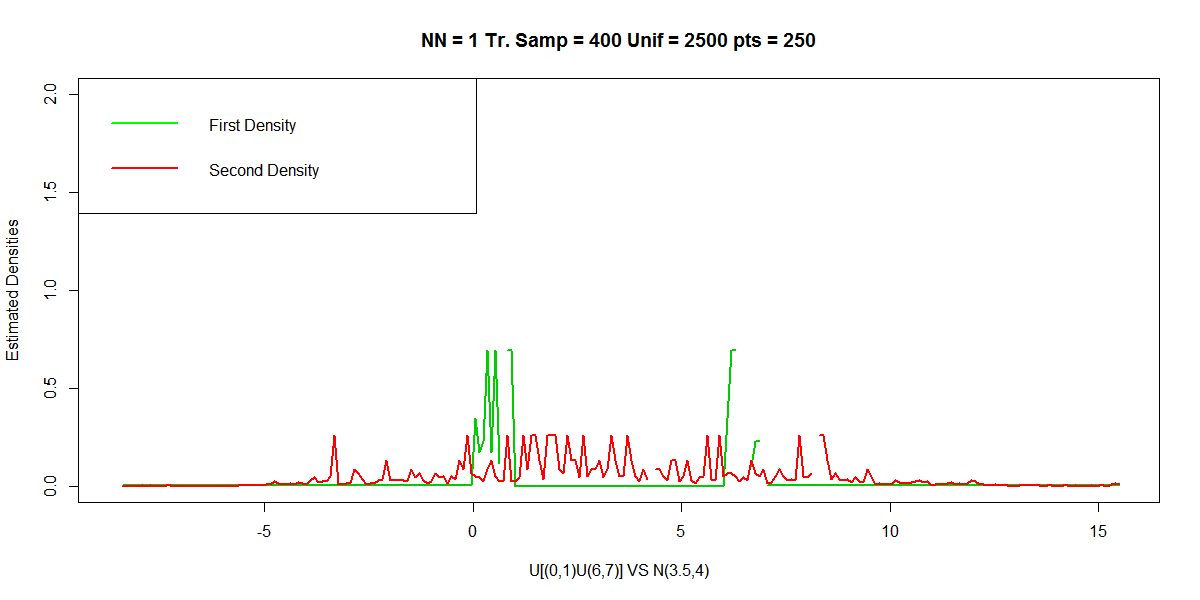}
  \caption{mUN-TS400-NN1}
  \label{mUN-TS400-NN1}
\end{subfigure}
\begin{subfigure}{.45\textwidth}
  \centering
  \includegraphics[width=.9\linewidth]{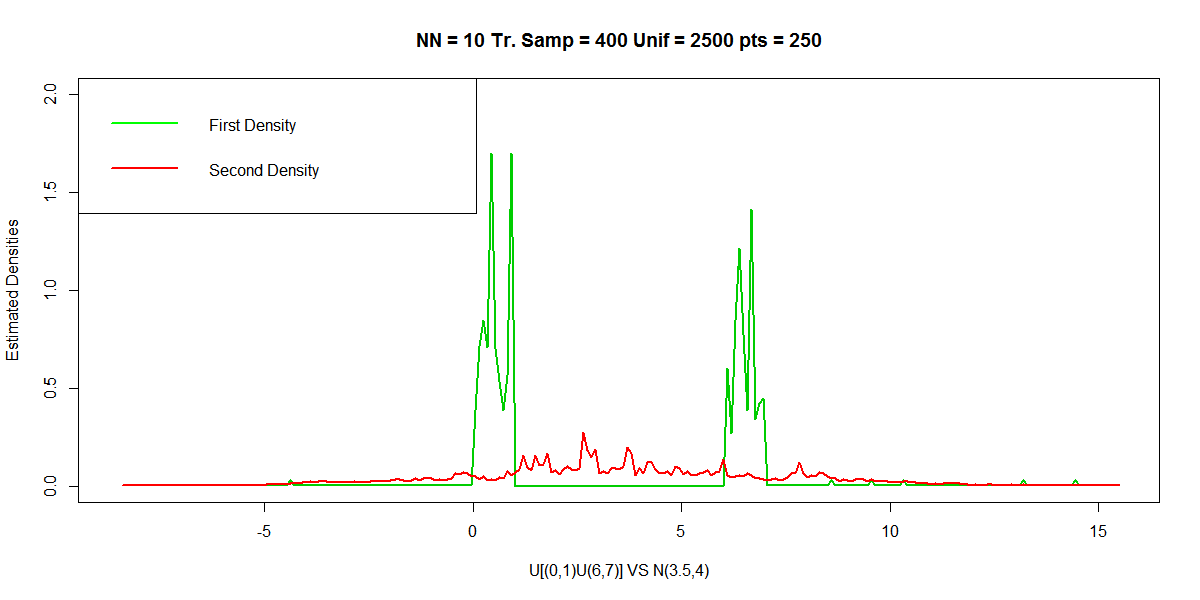}
  \caption{mUN-TS400-NN10}
  \label{mUN-TS400-NN10}
\end{subfigure}
\begin{subfigure}{.45\textwidth}
  \centering
  \includegraphics[width=.9\linewidth]{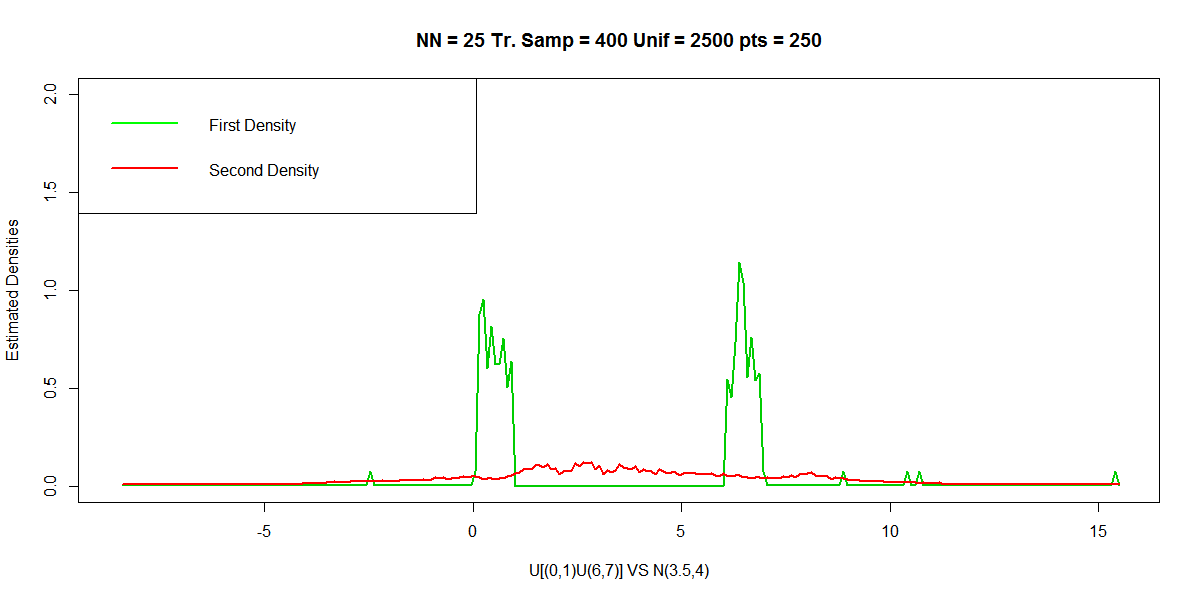}
  \caption{mUN-TS400-NN25}
  \label{mUN-TS400-NN25}
\end{subfigure}
\caption{Plots of densities for case 1 with 200 training samples and with 400 training samples }
\label{mUN}
\end{figure} 
 
  In case 1 we have taken the competing densities as $X_1\sim U[(0,1)\cup(6,7)]$ vs $X_2\sim N(3.5,4)$ . As we can see from the table~\ref{Case 1} our method easily beats the standard techniques like the Linear Discriminant Analysis, Quadratic Discriminant Analysis, Kernel Density Estimate using Gaussian Kernel and Kernel Density Estimate using Epanechnikov's Kernel. In this case our method performs significantly better than the existing methods. Here we can see that the number of observations originally coming from the second distribution, misclassified as coming from the first distribution is larger compared to the other type of misclassification.  Here LDA , QDA and Kernel based Bayes classification methods fails to compete with this method of classification. In this case we have taken the test sample to be of size 200 (100 from each class) and the training sample size is taken to be 200 and 400 from each class. Here as 200 is already moderately large so increasing the training sample does not significantly affect the misclassification rate in this case. Number of uniforms taken to estimate the Voronoi area of the cell is 1000 in each case. The two uniforms taken to estimate the Voronoi areas in this case are $U_1\sim U(-2,9)$ and $U_2 \sim(-8.5,15.5)$. While plotting number of uniforms taken is 2500. For classification taking 1000 uniforms produces good results.
  \begin{table}[!htbp]
\centering
\caption{Simulation Result for Case 1}
\label{Case 1}
\resizebox{\textwidth}{!}{%
\begin{tabular}{|c|c|c|c|c|c|c|c|}
\hline
\textbf{\begin{tabular}[c]{@{}c@{}}Competing\\   Distributions\end{tabular}}    & \textbf{\# of Unifroms} & \textbf{Training Sample} & \textbf{Method} & \textbf{-1} & \textbf{0} & \textbf{1} & \textbf{Misclass. Prob.} \\ \hline
\begin{tabular}[c]{@{}c@{}}Unif {[}(0,1)U(6,7){]} vs \\ N(3.5, 4)\end{tabular} & 1000                    & 200                      & NN1             & 8           & 174        & 18         & 0.13                     \\ \hline
                                                                                &                         &                          & NN10            & 9           & 171        & 20         & 0.145                    \\ \hline
                                                                                &                         &                          & NN25            & 5           & 176        & 19         & 0.12                     \\ \hline
                                                                                &                         &                          & LDA             & 44          & 101        & 55         & 0.495                    \\ \hline
                                                                                &                         &                          & QDA             & 31          & 110        & 59         & 0.45                     \\ \hline
                                                                                &                         &                          & G-Ker           & 0           & 137        & 63         & 0.315                    \\ \hline
                                                                                &                         &                          & E-Ker           & 0           & 128        & 72         & 0.36                     \\ \hline
                                                                                &                         & 400                      & NN1             & 20          & 166        & 14         & 0.17                     \\ \hline
                                                                                &                         &                          & NN10            & 5           & 176        & 19         & 0.12                     \\ \hline
                                                                                &                         &                          & NN25            & 6           & 175        & 19         & 0.125                    \\ \hline
                                                                                &                         &                          & LDA             & 44          & 100        & 56         & 0.5                      \\ \hline
                                                                                &                         &                          & QDA             & 2           & 139        & 59         & 0.305                    \\ \hline
                                                                                &                         &                          & G-Ker           & 0           & 134        & 66         & 0.33                     \\ \hline
                                                                                &                         &                          & E-Ker           & 0           & 133        & 67         & 0.335                    \\ \hline
\end{tabular}
}
\end{table}
  
\subsection*{Case 2}

\begin{figure}
\centering
\begin{subfigure}{.45\textwidth}
  \centering
  \includegraphics[width=.9\linewidth]{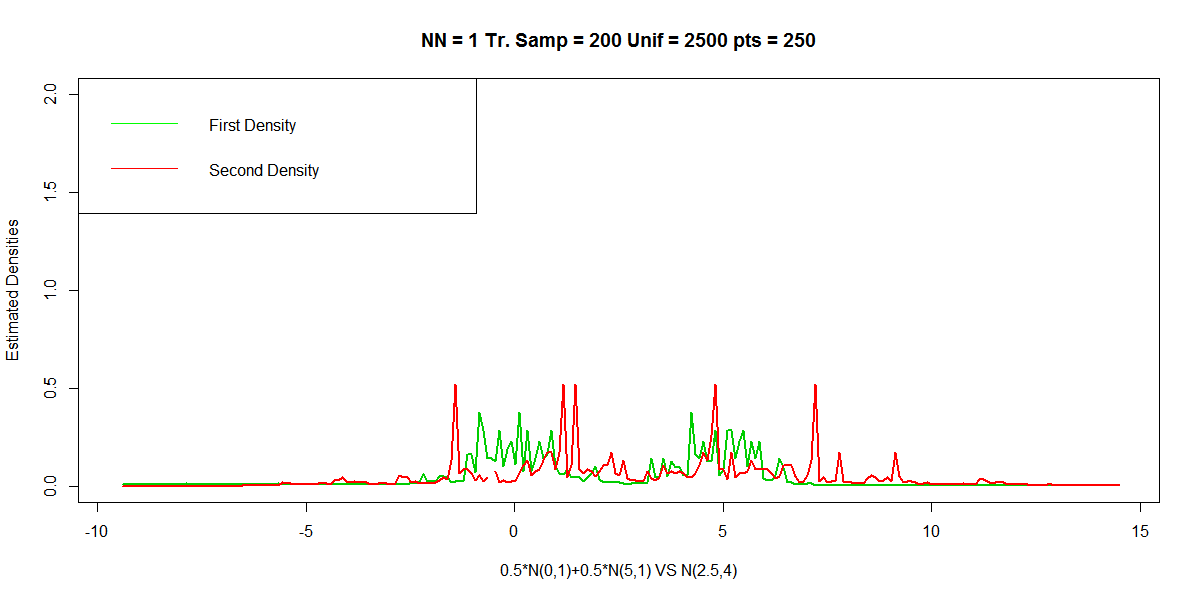}
  \caption{mNN-TS200-NN1}
  \label{mNN-TS200-NN1}
\end{subfigure}
\begin{subfigure}{.45\textwidth}
  \centering
  \includegraphics[width=.9\linewidth]{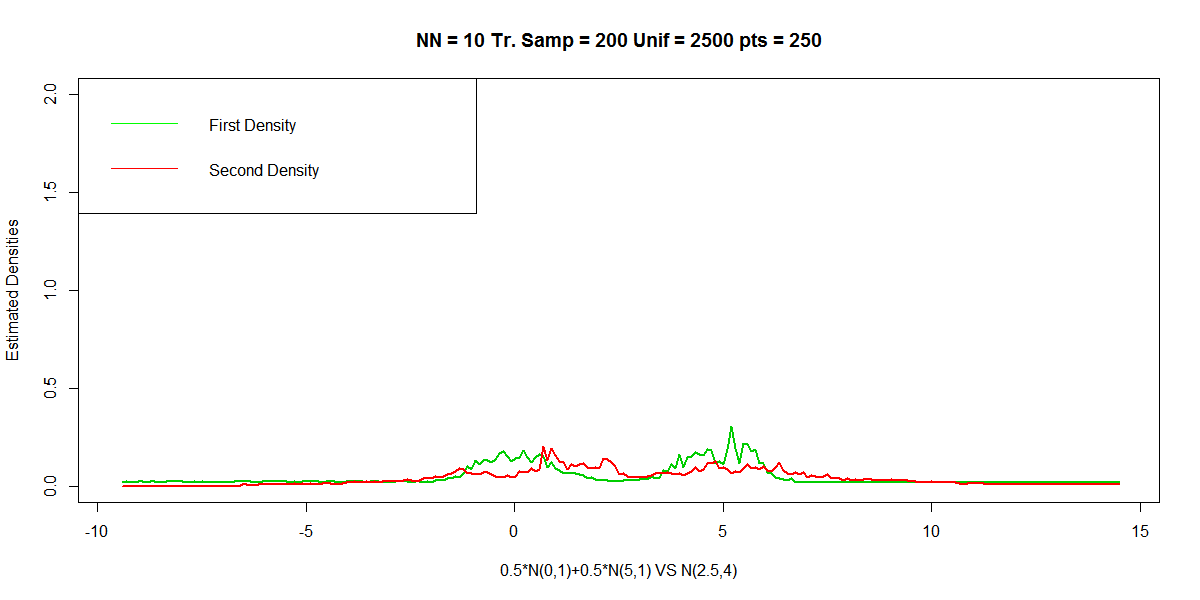}
  \caption{mNN-TS200-NN10}
  \label{mNN-TS200-NN10}
\end{subfigure}
\begin{subfigure}{.45\textwidth}
  \centering
  \includegraphics[width=.9\linewidth]{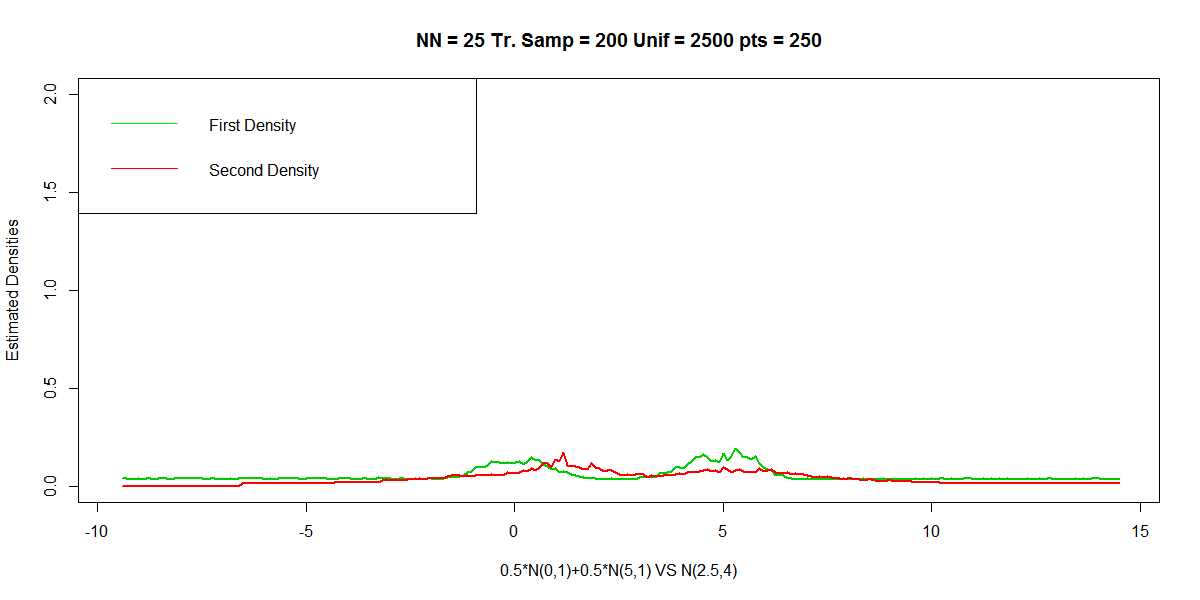}
  \caption{mNN-TS200-NN25}
  \label{mNN-TS200-NN25}
\end{subfigure}
\begin{subfigure}{.45\textwidth}
  \centering
  \includegraphics[width=.9\linewidth]{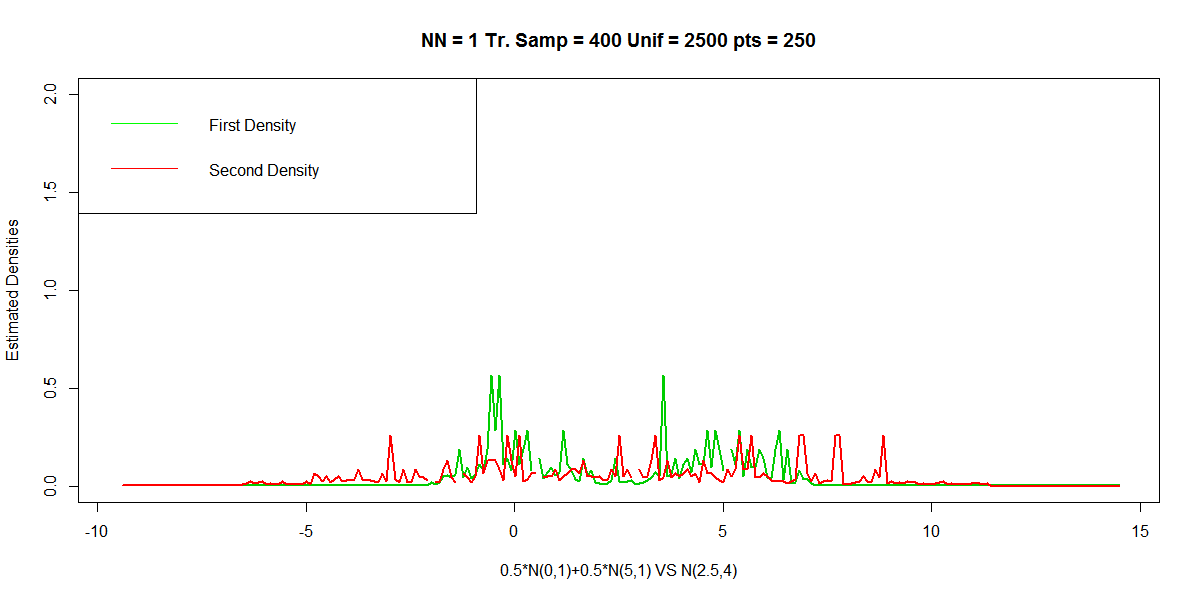}
  \caption{mNN-TS400-NN1}
  \label{mNN-TS400-NN1}
\end{subfigure}
\begin{subfigure}{.45\textwidth}
  \centering
  \includegraphics[width=.9\linewidth]{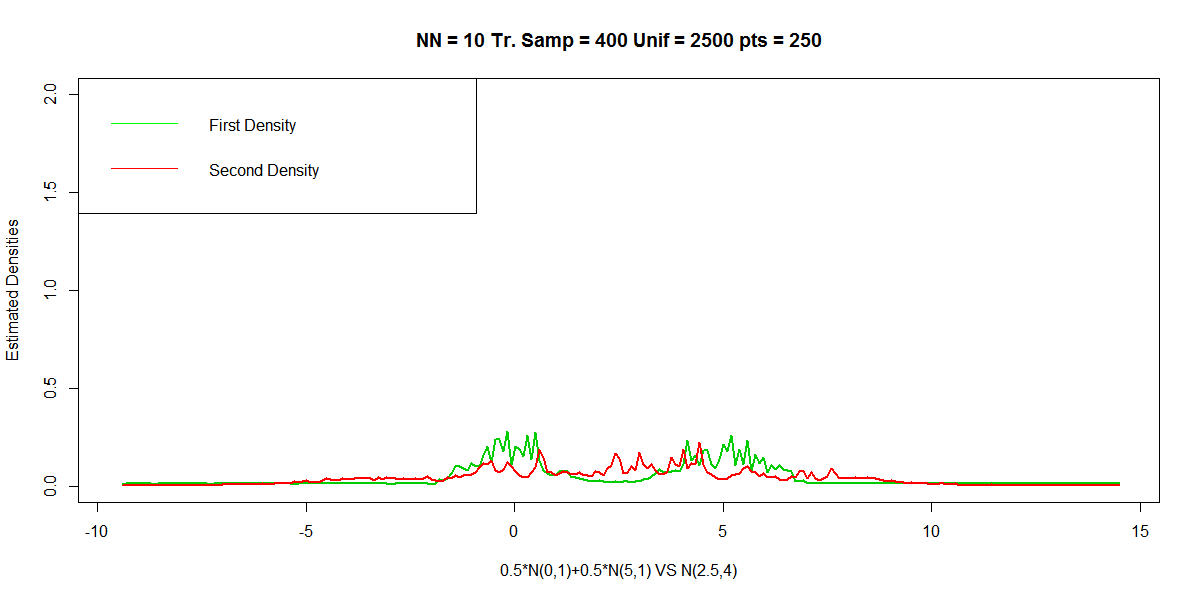}
  \caption{mNN-TS400-NN10}
  \label{mNN-TS400-NN10}
\end{subfigure}
\begin{subfigure}{.45\textwidth}
  \centering
  \includegraphics[width=.9\linewidth]{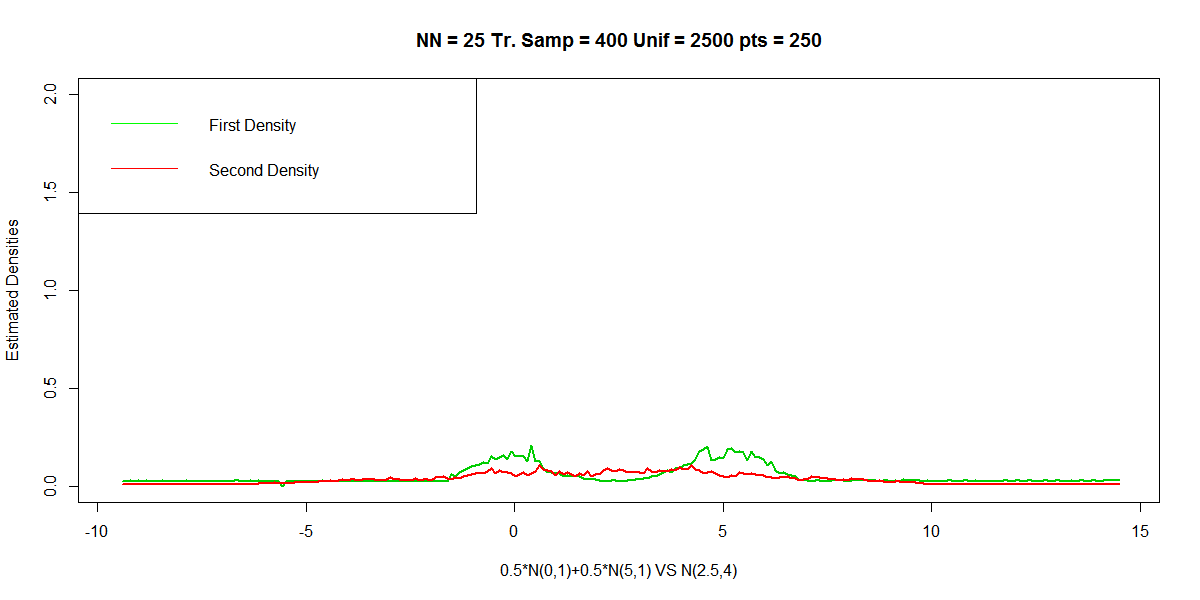}
  \caption{mNN-TS400-NN25}
  \label{mNN-TS400-NN25}
\end{subfigure}
\caption{Plots of densities for case 2 with 200 training samples and with 400 training samples }
\label{mNN}
\end{figure}

 In this case we have taken the two competing distributions to be $X_1 \sim \frac{1}{2}N(0,1)+\frac{1}{2}N(5,1)$ a mixture distribution of two normals vs $X_2\sim N(2.5,4)$. From the table~\ref{Case 2} we can see that our method perform significantly better than Linear Discriminant Analysis and performs at least as good as the Quadratic Discriminant Analysis. The Gaussian and Epanechnikov's Kernel Density Estimate based classification performs slightly better but not significantly better than our method. In this case also the observations originally coming from the second distribution classified as coming from the first distribution is larger than the other kind of misclassification in most of the cases. the uniforms taken to estimate the Voronoy area of the corresponding cells are $U_1 \sim U(-3,8)$ and $U_2 \sim U(-9.5,14.5)$. The training sample size is taken to be 200 from each and 400 from each. The test sample size is same as above. The number of uniforms taken to estimate the density is 1000 in some of the cases and 2500 in others. Increasing the number of uniforms slightly improves the misclassification and performs equivalent to the Gaussian and Epanechnikov's kernel.
\begin{table}[!htbp]
\centering
\caption{Simulation for Case 2}
\label{Case 2}
\resizebox{\textwidth}{!}{%
\begin{tabular}{|c|c|c|c|c|c|c|c|}
\hline
\textbf{\begin{tabular}[c]{@{}c@{}}Competing\\   Distributions\end{tabular}} & \textbf{\# of Unifroms} & \textbf{Training Sample} & \textbf{Method} & \textbf{-1} & \textbf{0} & \textbf{1} & \textbf{Misclass. Prob.} \\ \hline
\begin{tabular}[c]{@{}c@{}}$\frac{1}{2}$N(0,1)+$\frac{1}{2}$N(5,1)  \\vs  N(2.5,4)\end{tabular} & 1000 & 200 & NN1 & 42 & 119 & 39 & 0.405 \\ \hline
 &  &  & NN10 & 17 & 129 & 54 & 0.355 \\ \hline
 &  &  & NN25 & 17 & 123 & 60 & 0.385 \\ \hline
 & 2500 & & NN1 & 37 & 122 & 41 & 0.39 \\ \hline
 &  &  & NN10 & 30 & 128 & 42 & 0.36 \\ \hline
 &  &  & NN25 & 25 & 119 & 56 & 0.405 \\ \hline 
 &  &  & LDA & 48 & 99 & 53 & 0.505 \\ \hline
 &  &  & QDA & 22 & 119 & 59 & 0.405 \\ \hline
 &  &  & G-Ker & 12 & 136 & 52 & 0.32 \\ \hline
 &  &  & E-Ker & 14 & 127 & 59 & 0.365 \\ \hline
 & 1000  & 400 & NN1 & 34 & 123 & 43 & 0.385 \\ \hline
 &  &  & NN10 & 36 & 121 & 43 & 0.395 \\ \hline
 &  &  & NN25 & 20 & 120 & 60 & 0.4 \\ \hline
 & 2500 & & NN1 & 36 & 128 & 36 & 0.36 \\ \hline
 &  &  & NN10 & 27 & 127 & 46 & 0.365 \\ \hline
 &  &  & NN25 & 23 & 121 & 56 & 0.395 \\ \hline  
 &  &  & LDA & 52 & 98 & 50 & 0.51 \\ \hline
 &  &  & QDA & 28 & 119 & 53 & 0.405 \\ \hline
 &  &  & G-Ker & 19 & 129 & 52 & 0.355 \\ \hline
 &  &  & E-Ker & 10 & 130 & 60 & 0.35 \\ \hline
\end{tabular}
}
\end{table}
  
 \subsection*{Case 3} 
 
\begin{figure}
\centering
\begin{subfigure}{.45\textwidth}
  \centering
  \includegraphics[width=.9\linewidth]{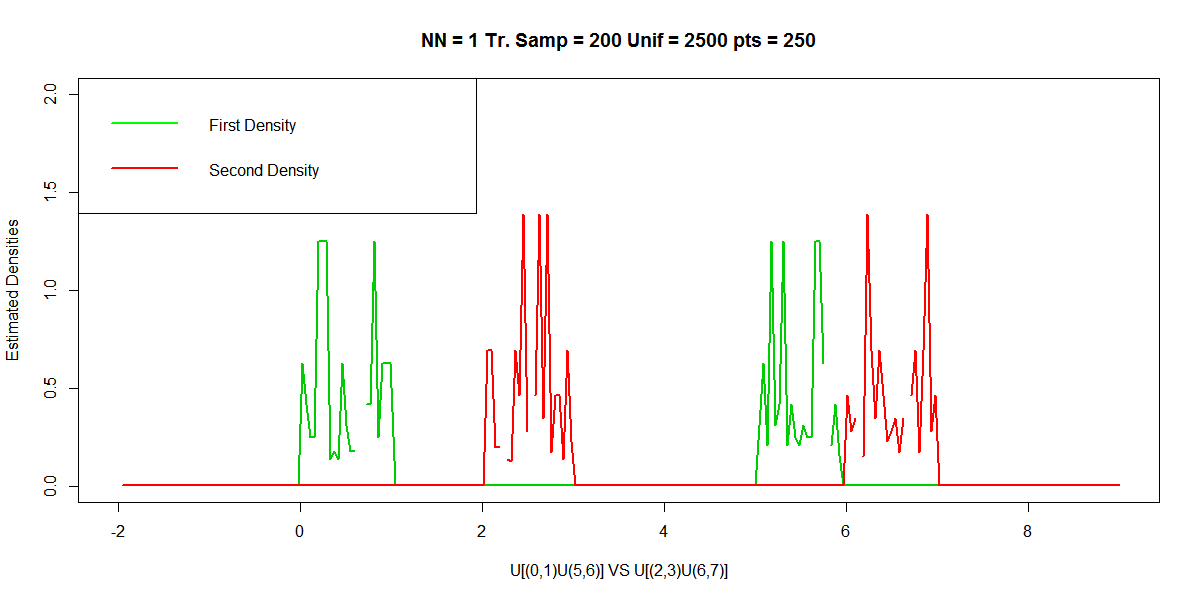}
  \caption{mUmU-TS200-NN1}
  \label{mUmU-TS200-NN1}
\end{subfigure}
\begin{subfigure}{.45\textwidth}
  \centering
  \includegraphics[width=.9\linewidth]{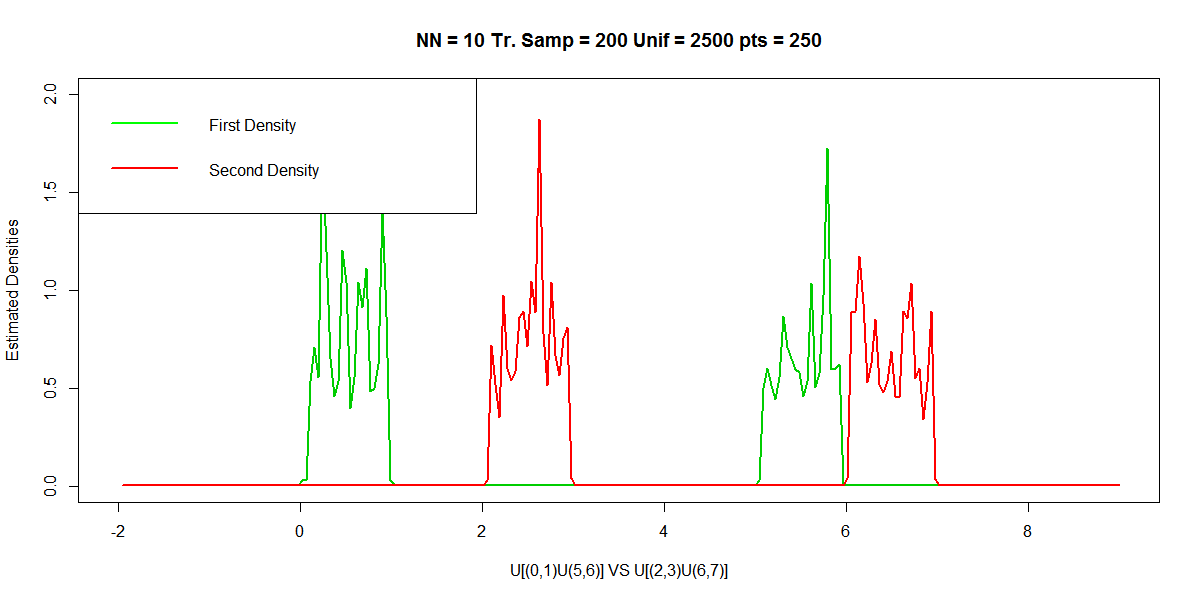}
  \caption{mUmU-TS200-NN10}
  \label{mUmU-TS200-NN10}
\end{subfigure}
\begin{subfigure}{.45\textwidth}
  \centering
  \includegraphics[width=.9\linewidth]{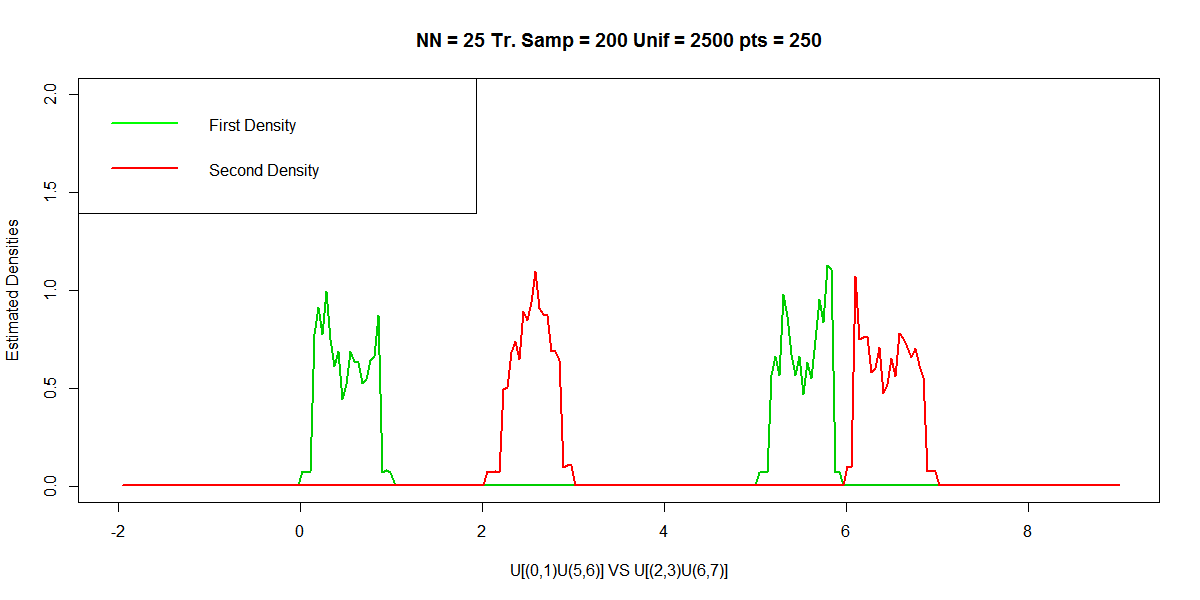}
  \caption{mUmU-TS200-NN25}
  \label{mUmU-TS200-NN25}
\end{subfigure}
\begin{subfigure}{.45\textwidth}
  \centering
  \includegraphics[width=.9\linewidth]{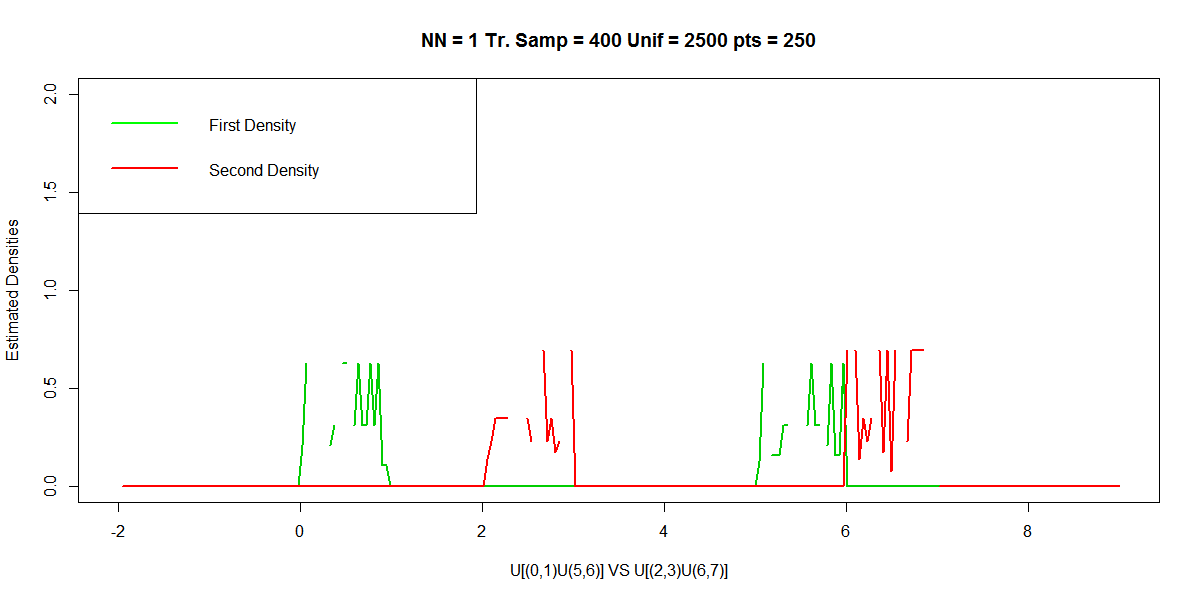}
  \caption{mUmU-TS400-NN1}
  \label{mUmU-TS400-NN1}
\end{subfigure}
\begin{subfigure}{.45\textwidth}
  \centering
  \includegraphics[width=.9\linewidth]{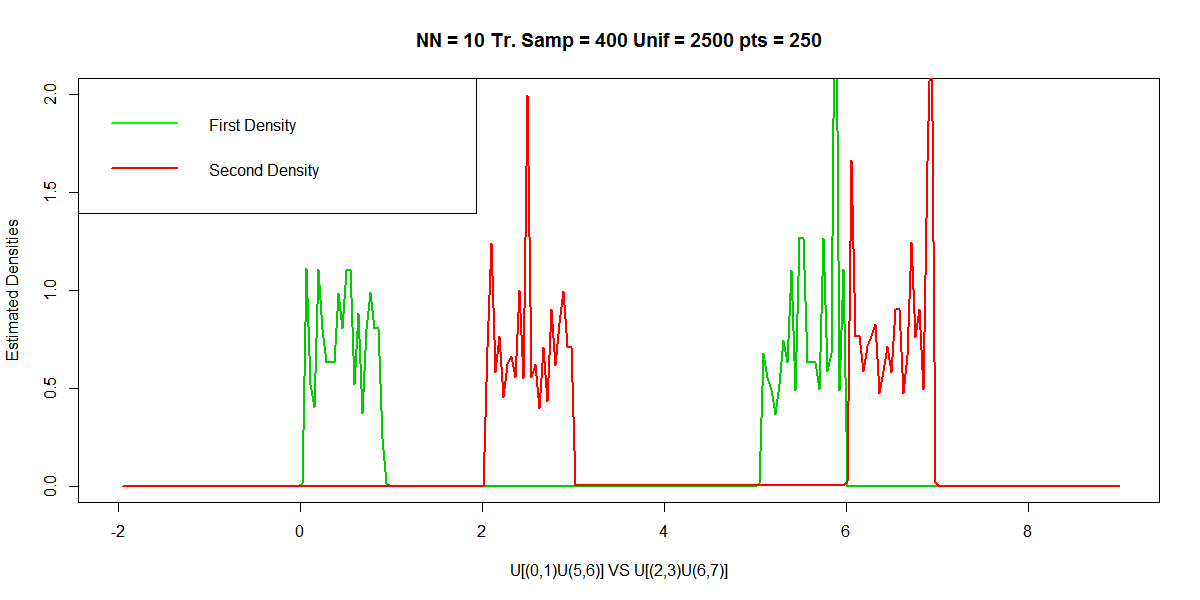}
  \caption{mUmU-TS400-NN10}
  \label{mUmU-TS400-NN10}
\end{subfigure}
\begin{subfigure}{.45\textwidth}
  \centering
  \includegraphics[width=.9\linewidth]{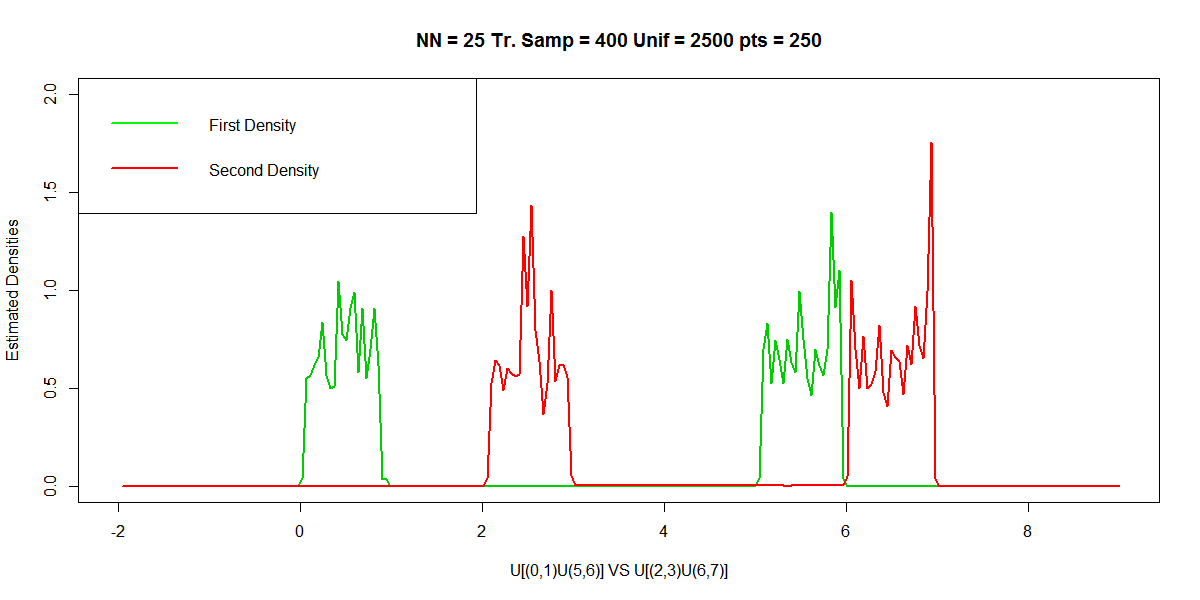}
  \caption{mUmU-TS400-NN25}
  \label{mUmU-TS400-NN25}
\end{subfigure}
\caption{Plots of densities for case 3 with 200 training samples and with 400 training samples }
\label{mUmU}
\end{figure}

   In this case we have taken the two competing distributions as $X_1 \sim U[(0,1)\cup(5,6)] $ and $X_2 \sim  U[(2,3)\cup(6,7)]$. From table~\ref{Case 3} we can see that Linear Discriminant Analysis and Quadratic Discriminant Analysis performs extremely poor with approximately $45\%$ of misclassification rate. On the contrary our method performs extremely well with less than $2.5\%$ of misclassification rate. The Kernel Density Estimate based Bayes classifier also produces similar result for using the Gaussian kernel and also the Epanechnikov's Kernel. So in this case our method conclusively beats the LDA and QDA methods and is equivalent to the Kernel Density Estimation method. Increase in training sample size does not effect significantly in the misclassification rate in this classification problem. The uniforms taken in this case are $U_1 \sim (-2,8)$ and $U_2 \sim U(0,9)$.  here also the training sample sizes are taken to be 200 and 400 in each of the cases and the test sample size is 200 such that 100 are from either of the distributions. In this case we do not need to increase the number of uniforms because it is already performing good.
  \begin{table}[!htbp]
\centering
\caption{Simulation result for Case 3}
\label{Case 3}
\resizebox{\textwidth}{!}{%
\begin{tabular}{|c|c|c|c|c|c|c|c|}
\hline
\textbf{\begin{tabular}[c]{@{}c@{}}Competing\\   Distributions\end{tabular}} & \textbf{\# of Unifroms} & \textbf{Training Sample} & \textbf{Method} & \textbf{-1} & \textbf{0} & \textbf{1} & \textbf{Misclass. Prob.} \\ \hline
\begin{tabular}[c]{@{}c@{}}U{[}(0,1)U(5,6){]}\\   vs  U{[}(2,3)U(6,7){]}\end{tabular} & 1000 & 200 & NN1 & 2 & 197 & 1 & 0.015 \\ \hline
 &  &  & NN10 & 2 & 196 & 2 & 0.02 \\ \hline
 &  &  & NN25 & 3 & 197 & 0 & 0.015 \\ \hline
 &  &  & LDA & 47 & 110 & 43 & 0.45 \\ \hline
 &  &  & QDA & 47 & 110 & 43 & 0.45 \\ \hline
 &  &  & G-Ker & 1 & 199 & 0 & 0.005 \\ \hline
 &  &  & E-Ker & 4 & 196 & 0 & 0.02 \\ \hline
 &  & 400 & NN1 & 2 & 198 & 0 & 0.01 \\ \hline
 &  &  & NN10 & 3 & 197 & 0 & 0.015 \\ \hline
 &  &  & NN25 & 2 & 198 & 0 & 0.01 \\ \hline
 &  &  & LDA & 47 & 110 & 43 & 0.45 \\ \hline
 &  &  & QDA & 47 & 110 & 43 & 0.45 \\ \hline
 &  &  & G-Ker & 3 & 197 & 0 & 0.015 \\ \hline
 &  &  & E-Ker & 4 & 196 & 0 & 0.02 \\ \hline
\end{tabular}
}
\end{table}

\subsection*{Case 4}

\begin{figure}
\centering
\begin{subfigure}{.45\textwidth}
  \centering
  \includegraphics[width=.9\linewidth]{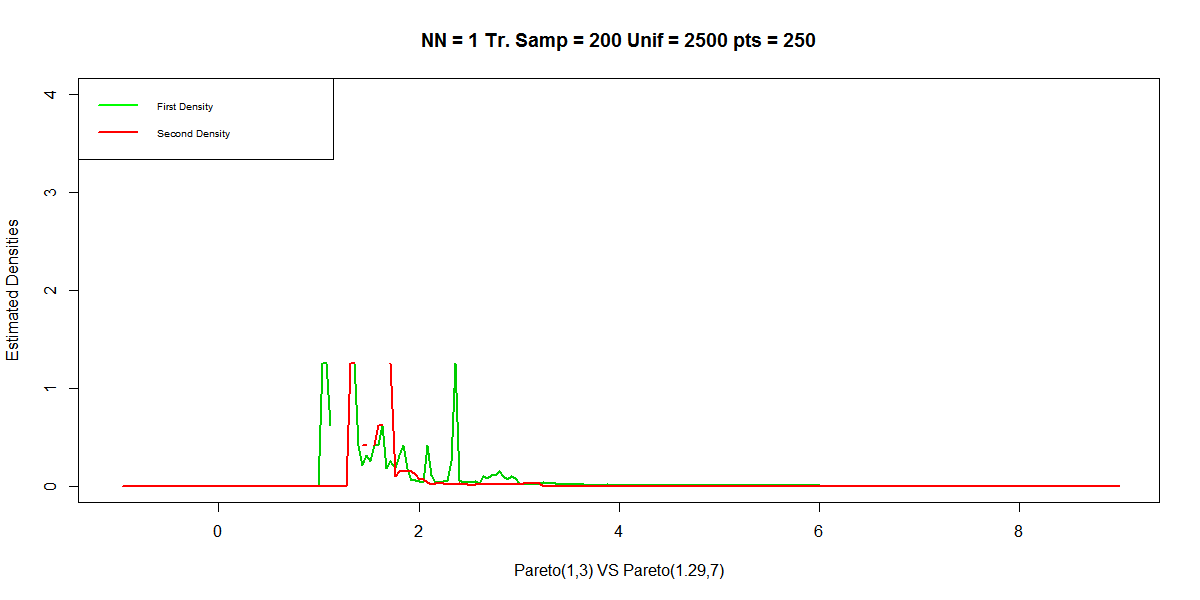}
  \caption{PP-TS200-NN1}
  \label{PP-TS200-NN1}
\end{subfigure}
\begin{subfigure}{.45\textwidth}
  \centering
  \includegraphics[width=.9\linewidth]{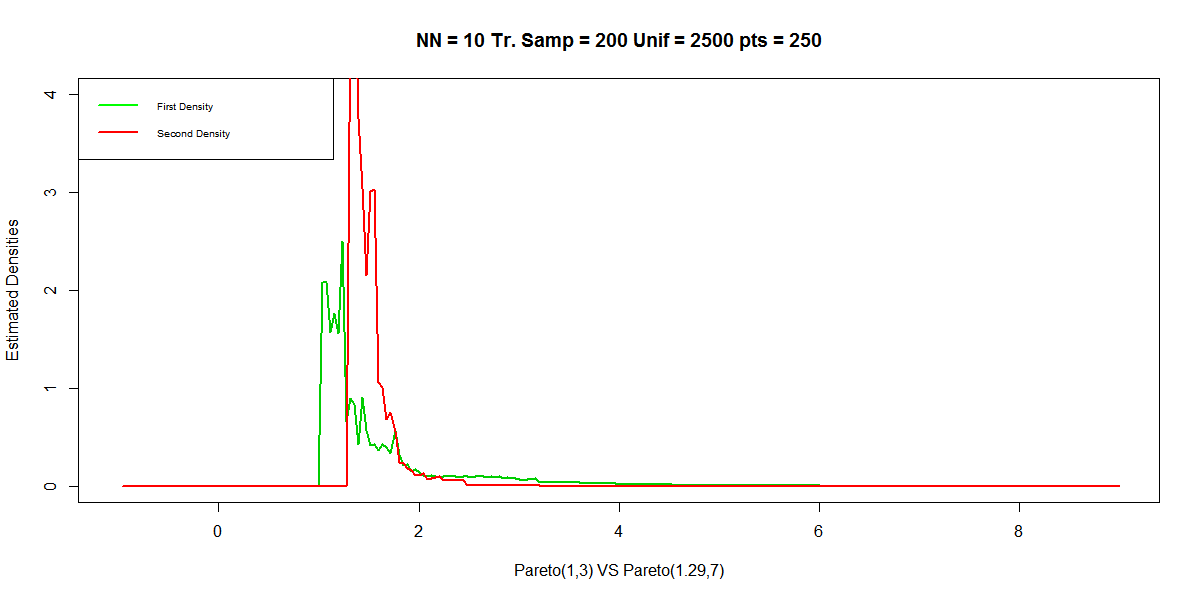}
  \caption{PP-TS200-NN10}
  \label{PP-TS200-NN10}
\end{subfigure}
\begin{subfigure}{.45\textwidth}
  \centering
  \includegraphics[width=.9\linewidth]{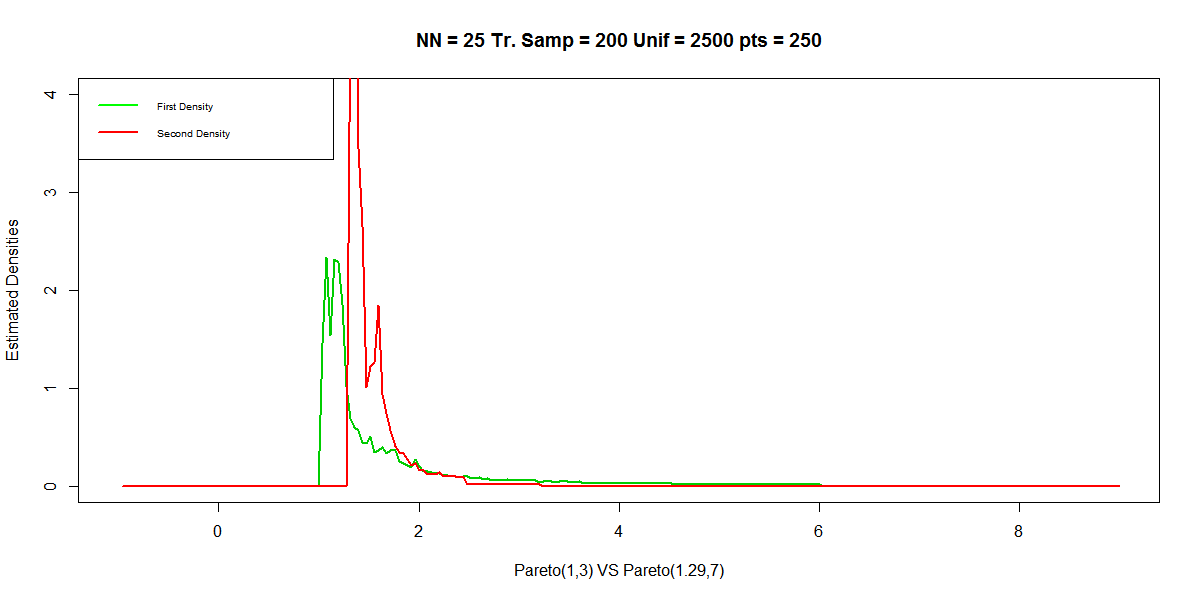}
  \caption{PP-TS200-NN25}
  \label{PP-TS200-NN25}
\end{subfigure}
\begin{subfigure}{.45\textwidth}
  \centering
  \includegraphics[width=.9\linewidth]{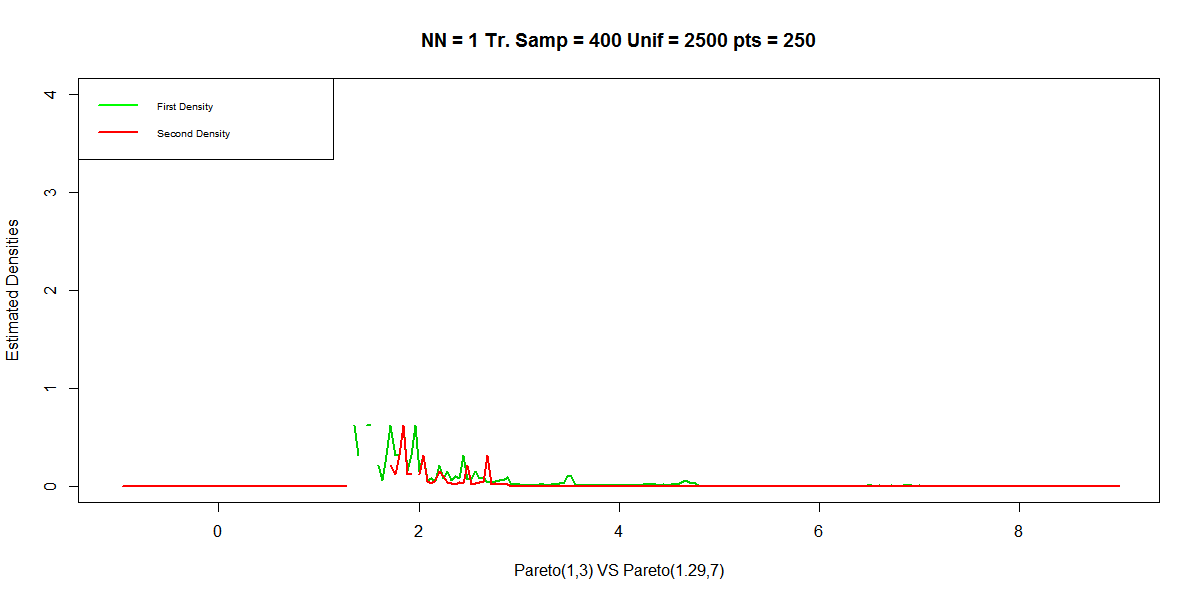}
  \caption{PP-TS400-NN1}
  \label{PP-TS400-NN1}
\end{subfigure}
\begin{subfigure}{.45\textwidth}
  \centering
  \includegraphics[width=.9\linewidth]{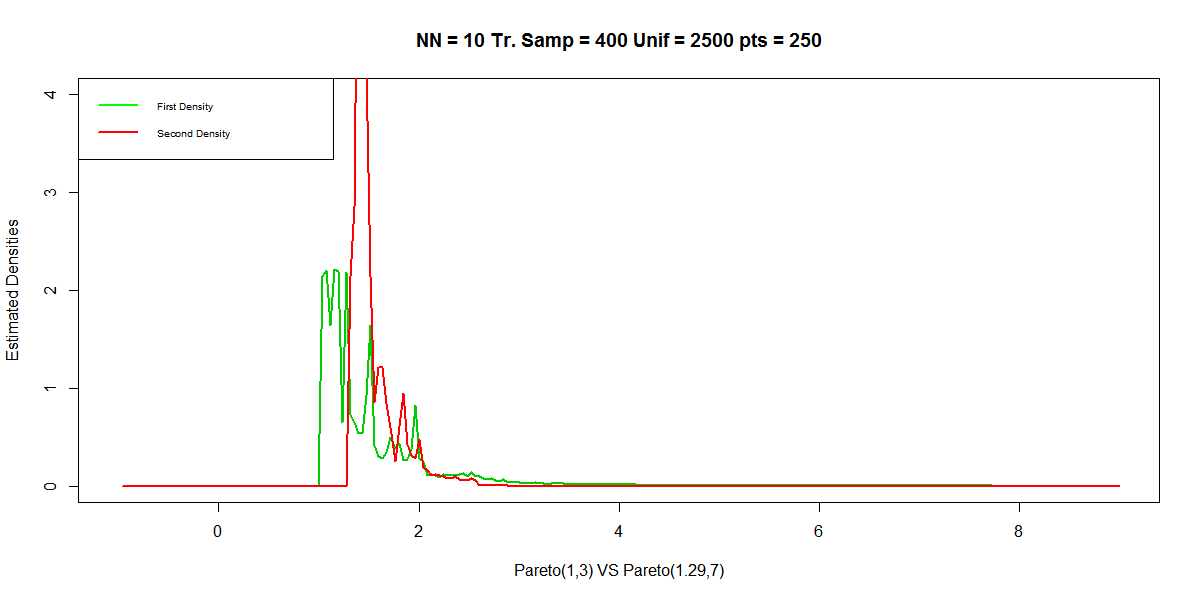}
  \caption{PP-TS400-NN10}
  \label{PP-TS400-NN10}
\end{subfigure}
\begin{subfigure}{.45\textwidth}
  \centering
  \includegraphics[width=.9\linewidth]{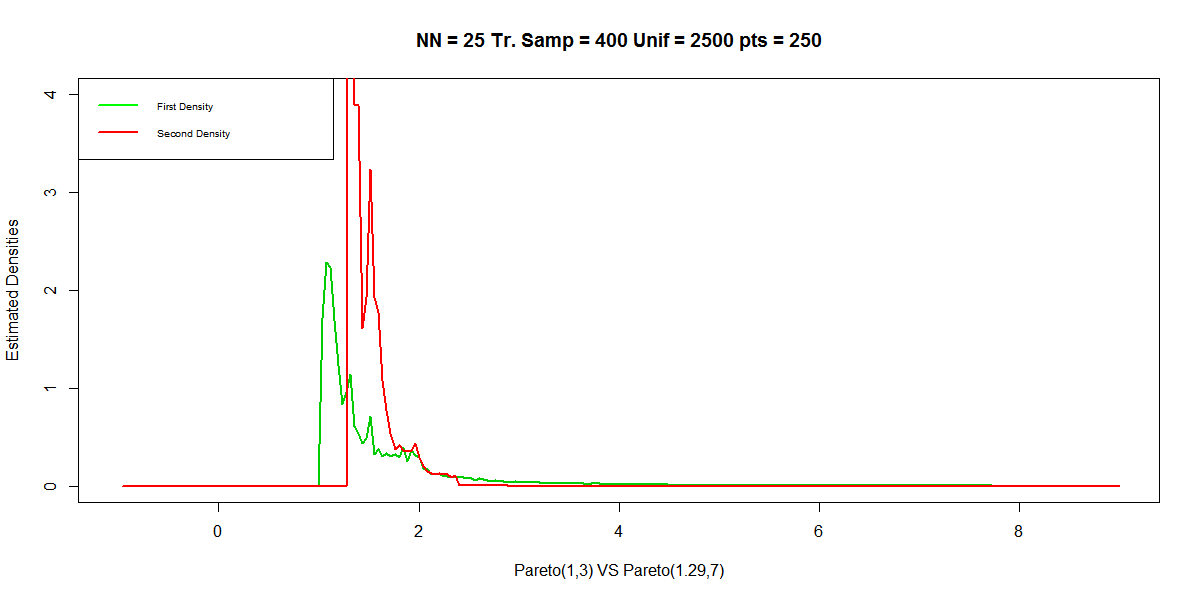}
  \caption{PP-TS400-NN25}
  \label{PP-TS400-NN25}
\end{subfigure}
\caption{Plots of densities for case 4 with 200 training samples and with 400 training samples }
\label{PP}
\end{figure}

 Her we have compared between two Pareto distributions which has the form as follows:
  \begin{align*}
  \text{pdf of the distribution} = f_{x_m,\alpha}(x) &= \frac{\alpha x_m^\alpha}{x^{\alpha+1}}  \qquad \text{for $x\geq x_m$}  
  \end{align*}
  The parameters of the distribution are $x_m$ and $\alpha$. We have taken $Pareto(1,3)$ and $Pareto(1.29,7)$. Here we can see some interesting facts. When we increase the number of nearest neighbours the misclassification rate decreases gradually. Ten nearest neighbour is better than single nearest neighbour and Twenty-five nearest neighbour is better than both one and ten nearest neighbour. The single nearest neighbour density estimation based classifier easily beats the Linear Discriminant Analysis and performs equivalent to Quadratic Discriminant Analysis. This is however not as good as the Gaussian and Epanechnikov's Kernel Density estimates based Bayes classifier. Increasing the number of nearest neighbour to 10 makes it better than LDA and QDA but its performance is still poorer than the Kernel Based Classifiers. When we increase the nearest neighbour to 25 then the classifier performs equivalent to the Gaussian and Epanechnikov's Kernel Density Estimate based Bayes Classifier. here we have taken the uniforms for estimating the Voronoi areas to be $U_1,U_2 \sim U(-1,9)$ both. Training and test sample size has been taken to be same as above.  
\begin{table}[!htbp]
\centering
\caption{Simulation for Case 4}
\label{Case 4}
\resizebox{\textwidth}{!}{%
\begin{tabular}{|c|c|c|c|c|c|c|c|}
\hline
\textbf{\begin{tabular}[c]{@{}c@{}}Competing\\   Distributions\end{tabular}} & \textbf{\# of Unifroms} & \textbf{Training Sample} & \textbf{Method} & \textbf{-1} & \textbf{0} & \textbf{1} & \textbf{Misclass. Prob.} \\ \hline
\begin{tabular}[c]{@{}c@{}}Pareto(1,3)\\ vs Pareto(1.29,7)\end{tabular} & 1000 & 200 & NN1 & 20 & 126 & 54 & 0.37 \\ \hline
 &  &  & NN10 & 37 & 144 & 19 & 0.28 \\ \hline
 &  &  & NN25 & 35 & 155 & 10 & 0.225 \\ \hline
 & 2500 & & NN1 & 19 & 150 & 31 & 0.25 \\ \hline
 &  &  & NN10 & 32 & 157 & 11 & 0.215 \\ \hline
 &  &  & NN25 & 35 & 158 & 7 & 0.21 \\ \hline 
 &  &  & LDA & 71 & 105 & 24 & 0.475 \\ \hline
 &  &  & QDA & 60 & 130 & 10 & 0.35 \\ \hline
 &  &  & G-Ker & 39 & 157 & 4 & 0.215 \\ \hline
 &  &  & E-Ker & 33 & 158 & 9 & 0.21 \\ \hline
 & 1000 & 400 & NN1 & 18 & 125 & 57 & 0.375 \\ \hline
 &  &  & NN10 & 32 & 146 & 22 & 0.27 \\ \hline
 &  &  & NN25 & 34 & 151 & 15 & 0.245 \\ \hline
& 2500 & & NN1 & 25 & 131 & 44 & 0.345 \\ \hline 
 & & & NN10 & 33 & 155 & 12 & 0.225 \\ \hline
 & & & NN25 & 34 & 158 & 8 & 0.21 \\ \hline
 &  &  & LDA & 33 & 97 & 70 & 0.515 \\ \hline
 &  &  & QDA & 64 & 128 & 8 & 0.36 \\ \hline
 &  &  & G-Ker & 45 & 153 & 2 & 0.235 \\ \hline
 &  &  & E-Ker & 42 & 152 & 6 & 0.24 \\ \hline
\end{tabular}
}
\end{table}

\subsection*{Case 5}

\begin{figure}
\centering
\begin{subfigure}{.45\textwidth}
  \centering
  \includegraphics[width=.9\linewidth]{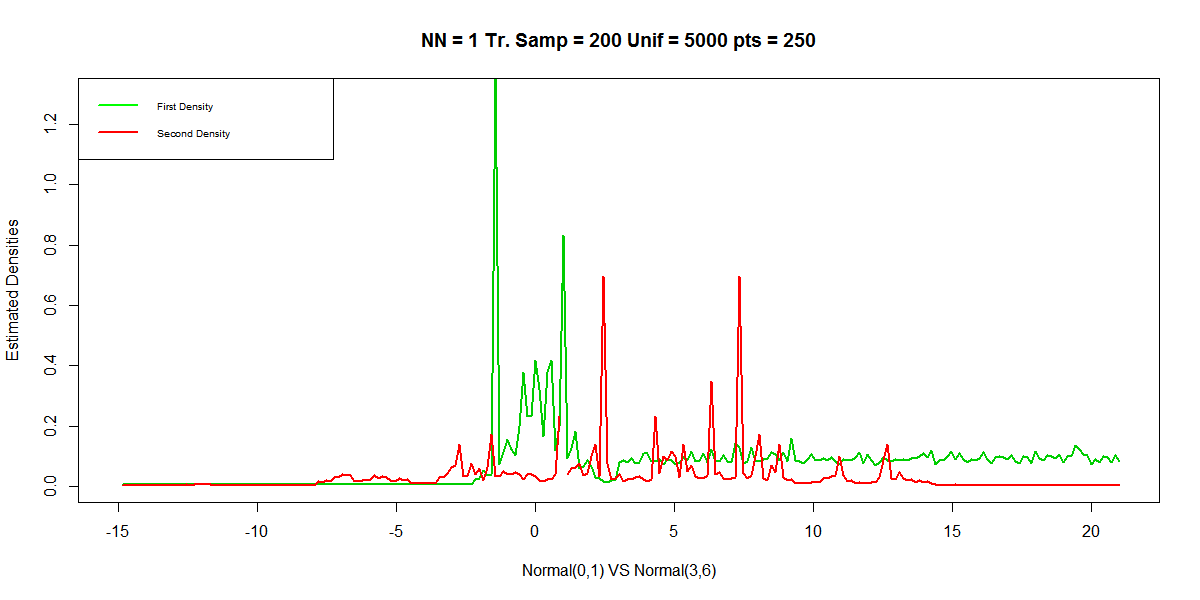}
  \caption{NN-TS200-NN1}
  \label{NN-TS200-NN1}
\end{subfigure}
\begin{subfigure}{.45\textwidth}
  \centering
  \includegraphics[width=.9\linewidth]{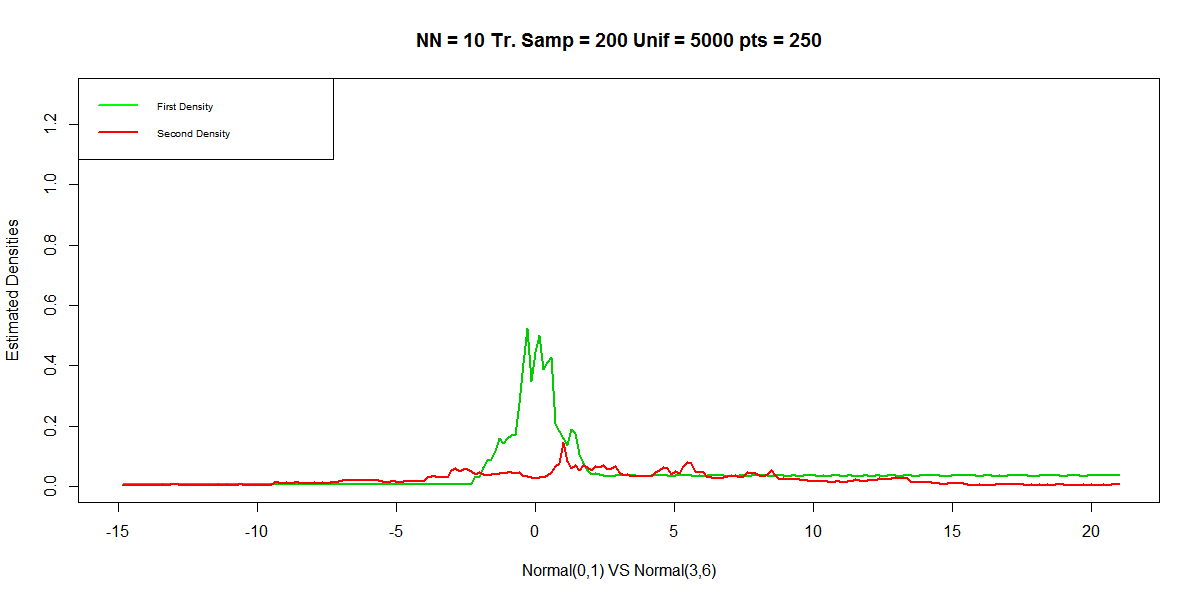}
  \caption{NN-TS200-NN10}
  \label{NN-TS200-NN10}
\end{subfigure}
\begin{subfigure}{.45\textwidth}
  \centering
  \includegraphics[width=.9\linewidth]{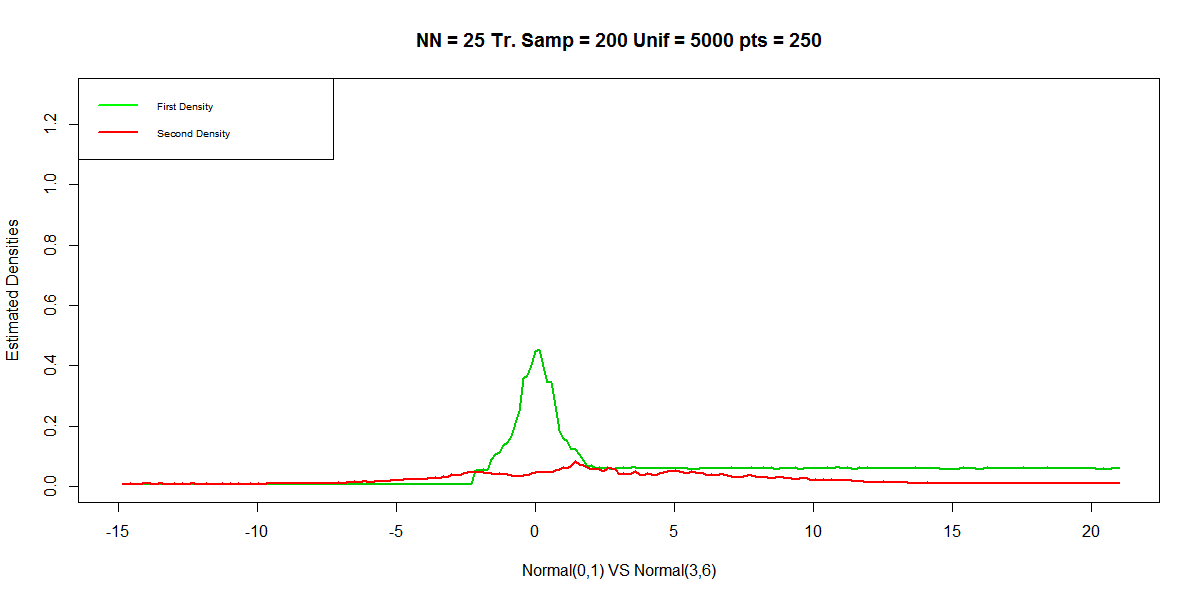}
  \caption{NN-TS200-NN25}
  \label{NN-TS200-NN25}
\end{subfigure}
\begin{subfigure}{.45\textwidth}
  \centering
  \includegraphics[width=.9\linewidth]{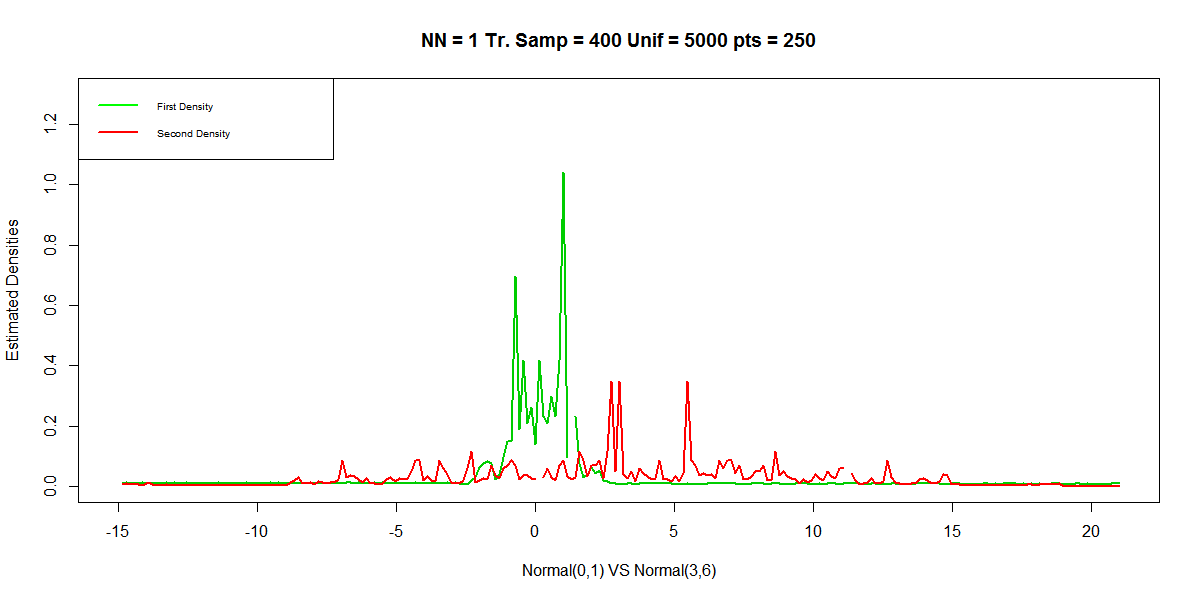}
  \caption{NN-TS400-NN1}
  \label{NN-TS400-NN1}
\end{subfigure}
\begin{subfigure}{.45\textwidth}
  \centering
  \includegraphics[width=.9\linewidth]{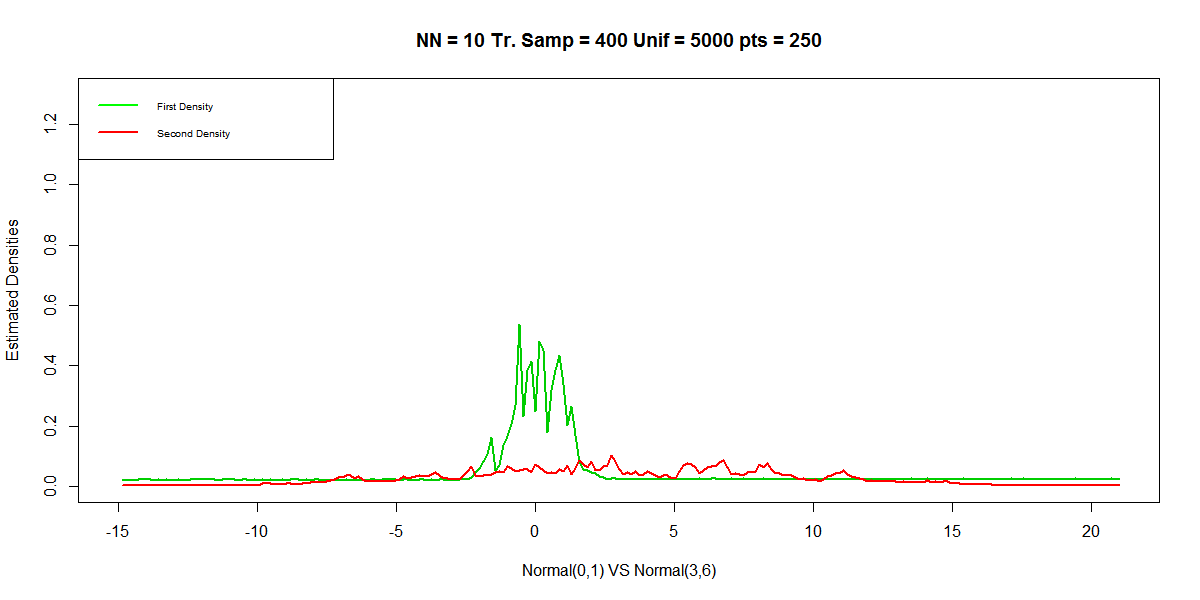}
  \caption{NN-TS400-NN10}
  \label{NN-TS400-NN10}
\end{subfigure}
\begin{subfigure}{.45\textwidth}
  \centering
  \includegraphics[width=.9\linewidth]{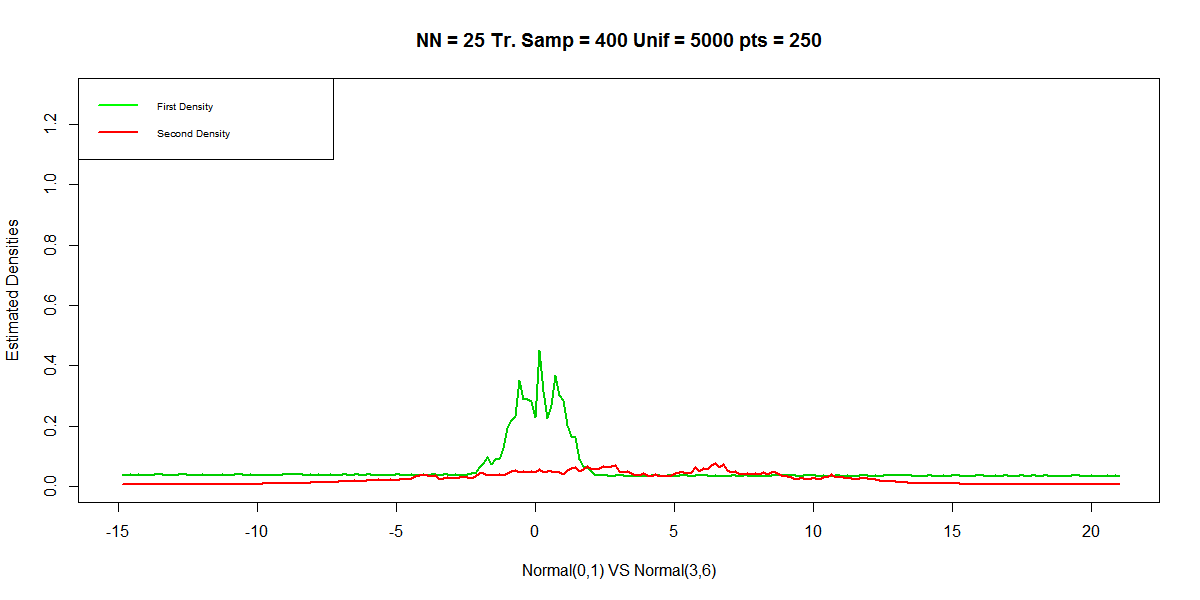}
  \caption{NN-TS400-NN25}
  \label{NN-TS400-NN25}
\end{subfigure}
\caption{Plots of densities for case 5 with 200 training samples and with 400 training samples }
\label{NN}
\end{figure}

 In this case we have taken the two completing distributions as $X_1\sim N(0,1)$ vs $X_2\sim N(3,6)$.  As in this case both the cases are normal and the training samples are coming from normal distribution so we expect that the QDA will give the best result. Here increasing the training sample does not increase the success rate in Bayes Classification. We can see here in our method  10 nearest neighbour method  performs better than 1 or 25 nearest neighbour Voronoi area based density estimation method. Here we can see increasing the training sample size improves the result. From this we can infer that 200 is not a large training sample in this case. With training sample size 400 our method with 10 nearest neighbour beats LDA, but still cannot beat QDA which is expected. Here we suspect that increasing the number of uniforms for estimating the Voronoi area will produce better result. Here we have taken the uniforms from $U_1 \sim U(-3,3)$ and $U_2 \sim U(-15,21)$. The number of training and test samples are same as above.  Here in this case QDA will produce the best result so, we have not taken into consideration of the Kernel Density Estimation based Bayes Classification Method.

\begin{table}[!htbp]
\centering
\caption{Simulation for Case 5}
\label{Case 5}
\resizebox{\textwidth}{!}{%
\begin{tabular}{|c|c|c|c|c|c|c|c|}
\hline
\textbf{\begin{tabular}[c]{@{}c@{}}Competing\\   Distributions\end{tabular}} & \textbf{\# of Unifroms} & \textbf{Training Sample} & \textbf{Method} & \textbf{-1} & \textbf{0} & \textbf{1} & \textbf{Misclass. Prob.} \\ \hline
\begin{tabular}[c]{@{}c@{}}N(0,1)\\   vs N(3,6)\end{tabular} & 1000 & 200 & NN1 & 0 & 100 & 100 & 0.5 \\ \hline
 &  &  & NN10 & 5 & 137 & 58 & 0.315 \\ \hline
 &  &  & NN25 & 3 & 123 & 74 & 0.385 \\ \hline
  & 5000 & & NN1 & 0 & 100 & 100 & 0.5 \\ \hline
 & & & NN10 & 10 & 128 & 62 & 0.36 \\ \hline
 & & & NN25 & 4 & 110 & 86 & 0.45 \\ \hline
 &  &  & LDA & 5 & 147 & 48 & 0.265 \\ \hline
 &  &  & QDA & 9 & 161 & 30 & 0.195 \\ \hline
 & 1000 & 400 & NN1 & 26 & 138 & 36 & 0.31 \\ \hline
 &  &  & NN10 & 8 & 148 & 44 & 0.26 \\ \hline
 &  &  & NN25 & 4 & 139 & 57 & 0.305 \\ \hline
 & 5000 & & NN1 & 10 & 155 & 35 & 0.225 \\ \hline
 & & & NN10 & 7 & 150 & 43 & 0.25 \\ \hline
 & & & NN25 & 3 & 134 & 63 & 0.33 \\ \hline
 &  &  & LDA & 8 & 145 & 47 & 0.275 \\ \hline
 &  &  & QDA & 6 & 161 & 33 & 0.195 \\ \hline
\end{tabular}%
}
\end{table}

\subsubsection*{Comment}
  
  As the two competing densities are normal, it is expected that QDA will perform much better compared to our method as QDA is optimal in case of normal distribution.

\subsection*{Case 6}



Here we have taken the first distribution as the union of two annulus i.e. we generated 2D observations with norm $r \in [(5,6)\cup(9,10)]$.  The second distribution is from a two dimensional normal distribution  $X_2\sim N(\utilde{0},36I_2)$. Here the uniforms to estimate the Voronoi area for the first distribution $U_1$ is generated from the annulus with radius between 4 and 11 : i.e. to cover the support. For the estimation of Voronoi cell area for estimating the second distribution we have taken uniform $U_2$ on the circle of radius 18. Here we have taken 1000 uniforms for estimation of densities. The training sample sizes are taken to be 200 and 400. increasing the training sample size improves the result significantly. For training sample size 400 we can see that LDA, QDA cannot perform equivalent to our method irrespective of the number of nearest neighbour. Our method is significantly better in this case.Here also the number of observation originally from second distribution classified as the first distribution is higher compared to the other type of misclassification.
\begin{table}[!htbp]
\centering
\caption{Simulation for Case 6}
\label{Case 6}
\resizebox{\textwidth}{!}{%
\begin{tabular}{|c|c|c|c|c|c|c|c|}
\hline
\begin{tabular}[c]{@{}c@{}}Competing\\   Distributions\end{tabular} & \# of Unifroms & Training Sample & Method & -1 & 0 & 1 & Misclass. Prob. \\ \hline
\begin{tabular}[c]{@{}c@{}}Case6  \end{tabular} & 1000 & 200 & NN1 & 22 & 114 & 64 & 0.43 \\ \hline
\begin{tabular}[c]{@{}c@{}}Unif with radius\\   $[(5,6)\cup(9,10)]$\end{tabular} &  &  & NN10 & 5 & 106 & 89 & 0.47 \\ \hline
$N(\utilde{0},36I_2)$ &  &  & NN25 & 1 & 106 & 93 & 0.47 \\ \hline
 &  &  & LDA & 44 & 105 & 51 & 0.475 \\ \hline
 &  &  & QDA & 39 & 114 & 47 & 0.43 \\ \hline
 &  & 400 & NN1 & 27 & 120 & 53 & 0.4 \\ \hline
 &  &  & NN10 & 8 & 110 & 82 & 0.45 \\ \hline
 &  &  & NN25 & 6 & 103 & 91 & 0.485 \\ \hline
 &  &  & LDA & 45 & 98 & 57 & 0.51 \\ \hline
 &  &  & QDA & 52 & 80 & 68 & 0.6 \\ \hline
\end{tabular}%
}
\end{table}

\subsubsection*{Comment}
 As the bivariate density considered here is not uni-modal, LDA and QDA perform worse than our methods when number of training sample is large. The performance in 200 samples in 2-dimension due to random factors LDA and QDA performs slightly good.

\subsection*{Case 7}
 Here we have taken the two competing distributions as uniform on a circle and another distribution on a circle.  For generating uniform from a circle we have first generated $X\sim N(\utilde{0},I_2)$  and then have taken $X_1=\frac{X}{||X||}$.
\[
\Sigma=
  \begin{bmatrix}
    1 & 0 \\
    0 & 4 
  \end{bmatrix}
\]
 Similarly for the second competing distribution we have taken $Y \sim N(\utilde{0},\Sigma)$ and then we have taken $X_2=\frac{Y}{||Y||}$. These are the competing distributions we have taken. As we can see from the table~\ref{Case 7} that our method irrespective of the number of nearest neighbour beats LDA surely and performs almost equivalent to QDA in this case. Increasing the training sample size improves the classification result. To estimate the Voronoi area we have taken uniforms from the circle in the same method as above: 1000 uniforms for estimating each of the densities. Training sample size is increased from 200 to 400 and the corresponding result has improved. We expect that increasing the number of uniforms and the number of points in training sample may improve the result further.

\begin{table}[!htbp]
\centering
\caption{Simulation for Case 7}
\label{Case 7}
\resizebox{\textwidth}{!}{%
\begin{tabular}{|c|c|c|c|c|c|c|c|}
\hline
\begin{tabular}[c]{@{}c@{}}Competing\\   Distributions\end{tabular} & \# of Unifroms & Training Sample & Method & -1 & 0 & 1 & Misclass. Prob. \\ \hline
$X \sim N(0,I) $ then  $X/||X||$ & 1000 & 200 & NN1 & 38 & 106 & 56 & 0.47 \\ \hline
$N(0,\Sigma)$ then  $X/||X||$ &  &  & NN10 & 51 & 99 & 50 & 0.505 \\ \hline

$\Sigma=
  \begin{bmatrix}
    1 & 0 \\
    0 & 4 
  \end{bmatrix}$
 &  &  & NN25 & 38 & 112 & 50 & 0.44 \\ \hline
 &  &  & LDA & 51 & 86 & 63 & 0.57 \\ \hline
 &  &  & QDA & 46 & 112 & 42 & 0.44 \\ \hline
 &  & 400 & NN1 & 37 & 113 & 50 & 0.435 \\ \hline
 &  &  & NN10 & 36 & 126 & 38 & 0.37 \\ \hline
 &  &  & NN25 & 42 & 121 & 37 & 0.395 \\ \hline
 &  &  & LDA & 51 & 110 & 39 & 0.45 \\ \hline
 &  &  & QDA & 39 & 124 & 37 & 0.38 \\ \hline
\end{tabular}%
}
\end{table}

  \subsubsection*{Comments}
   Our distribution on circle are derived from bivariate normal densities, which forces QDA and LDA to perform slightly good. There are many densities, trimodal or with more modes. Then it is expected that our method will beat LDA and QDA by far.

  \section*{Future Work}
  In typical manifolds and higher dimensions it is a very difficult problem to get the area of the Voronoi cells. Contrary to one dimension, unless we can estimate the area of the Voronoi cells, density estimation in this method will be a failure. But it is easy to see that all our methods done for one dimension ditto follows for manifolds and higher dimension as  far as the simulation is concerned. But to match the theoretical background these particular cases should be thought out. So simulation studies will be performed to see the behaviour and efficiency of this Voronoi Area Based Nearest Neighbour Density Estimation Method (VABNNDEM) in Bayes Classification Rule in higher dimensions. Initial simulations show fairly good misclassification rate when compared to the other methods of density estimation based classification techniques and standard ones. Instead of uniform we can use other distributions also which will provide a dominating probability measure. We have also used simulation to observe a good performance on distributions with ellipsoidal and spherical support. More simulation studies along with comparison with the existing methods will be performed in the next part in future. We also have plan to extend this method to weighted nearest neighbours with weights proportional to the distances. 
  
  \section*{Acknowledgement}
   We are grateful to Prof. Probal Chaudhuri\footnote{Indian Statistical Institute, Kolkata} for his helpful comments in ameliorating this work.

\end{document}